\documentclass[journal]{IEEEtran}

\usepackage{balance}
\usepackage{cite}
\usepackage{float}
\usepackage{multicol,multirow}
\usepackage{makecell,booktabs}
\usepackage{url}
\usepackage{hyperref}
\hypersetup{    
    colorlinks=true,
    linkcolor=black,
    anchorcolor=black,
    citecolor=black,
    filecolor=black,
    menucolor=black,
    runcolor=black,
    urlcolor=black,
}

\usepackage{subfigure}
\usepackage[utf8]{inputenc} 
\usepackage{graphicx}
\usepackage{amsfonts}
\usepackage{amssymb}
\usepackage{amsmath,mathtools}
\usepackage{array}

\usepackage{color}
\usepackage{svg}
\usepackage{bbm}
\usepackage{bm}
\usepackage{comment}
\usepackage{forest}
\usepackage{etoolbox}
\usepackage{xspace}
\usepackage[utf8]{inputenc} 

\usepackage{cleveref}
\usepackage{kotex}
\usepackage{lipsum}  
\usepackage{balance}
\usepackage{tabularx}
\usepackage{cite}
\usepackage{float}
\usepackage{multicol}
\usepackage{multirow}
\usepackage{makecell,booktabs}
\usepackage{graphicx}
\usepackage{subcaption}

\newcolumntype{L}[1]{>{\raggedright\let\newline\\\arraybackslash\hspace{0pt}}m{#1}}
\newcolumntype{C}[1]{>{\centering\let\newline\\\arraybackslash\hspace{0pt}}m{#1}}
\newcolumntype{R}[1]{>{\raggedleft\let\newline\\\arraybackslash\hspace{0pt}}m{#1}}

\begin{document}

\author{
    Hyun~Joong~Kim,~\IEEEmembership{Member,~IEEE,}
    and Jip~Kim,~\IEEEmembership{Member,~IEEE,}
}

\title{
    {Inertia-aware Unit Commitment and Remuneration Methods for Decarbonized Power System}
}
\maketitle

\begin{abstract}
High penetration of inverter-based resources (IBRs) reduces synchronous inertia and makes frequency stability harder to maintain in renewable-rich, low-inertia power systems. 
This paper develops an inertia-aware chance-constrained unit commitment (CC-UC) model that co-optimizes energy, reserves, and inertia, explicitly modeling time-coupled constraints of synchronous generators (SGs) and stochastic uncertainty in photovoltaic and wind generation and their inertia contributions, while co-optimizing inertia from storage. 
Within this framework, we implement four remuneration schemes for inertia—uplift, marginal pricing (MP), approximate convex hull pricing (aCHP), and average incremental pricing (AIP)—and use them for market settlement. 
Case studies on the IEEE 118-bus and a realistic Korean system show that incorporating inertia constraints improves frequency nadir and RoCoF and achieves the required inertia with fewer online SGs than a reliability-must-run (RMR) approach. 
Among the remuneration schemes, aCHP ensures revenue adequacy with the lowest uplift and provides the strongest price signal for inertia; MP requires the largest uplift because inertia is weakly valued in its price formation; and AIP yields intermediate performance, partially internalizing fixed costs into energy and reserve prices but still requiring non-negligible uplift. 
These results provide practical guidance for designing cost-reflective inertia markets in future low-inertia power systems.
\end{abstract}

\section{Introduction}
The penetration of renewable energy sources (RES) has rapidly increased following a global paradigm shift toward decarbonization. According to~\cite{RES_capacity}, the capacity of the photovoltaic generator (PV) and wind turbine (WT) increased from 40 GW and 181 GW in 2010 to 1,866 GW and 1,133 GW in 2024. However, this rapid penetration of RES leads to a reduction in the rotational inertia provided by synchronous generators (SG), raising concerns about frequency stability.

One promising approach is to leverage the capabilities of inverter-based resources (IBRs) to provide fast frequency response (FFR) or virtual inertia~\cite{zhong2010synchronverters}. These services can be delivered using control strategies that inject active power in response to a sudden power imbalance. According to IEEE standards~\cite{8332112}, FFR is defined to operate within 0.1 to 1 second after a disturbance reaches a predefined threshold, aiming to mitigate the frequency nadir. However, its response is slower than that of inertial response and may induce control interactions such as overshoot. Inertial response acts within 0.05 to 0.3 seconds, modulating power output in proportion to the rate of change of frequency (RoCoF) signal.

A significant reduction in system inertia can result in a very high RoCoF. A rapid RoCoF indicates a rapid frequency decline during contingency events, potentially disrupting under-frequency load shedding that relies on early-stage frequency detection and response. 
Such disruptions may lead to blackouts or frequency collapse. 
Furthermore, inertia influences the system to damp oscillations and to maintain RoCoF within acceptable limits, thus preserving signal stability through adequate inertial support~\cite{denholm2020inertia}.

\begin{table*}[t]
  \centering
  \footnotesize  
  \centering
  \vspace{-2.0mm}
    \captionsetup{justification=centering, labelsep=period, font=footnotesize, textfont=sc}
  \setlength{\tabcolsep}{3.5pt}
\centering
\caption{Comparison of Inertia-aware System Operation and Remuneration Policy }
\label{tab:inertia_comparison}
\begin{tabular}{L{2.3cm}|L{4.0cm}|L{3.5cm}|L{5.8cm}}
\toprule
\textbf{} & \textbf{System Operation} & \textbf{Available Resource} & \textbf{Remuneration Policy} \\
\midrule
ERCOT~\cite{ERCOT2024}& Reliability Must-Run & SG/SD & Contract\\
CAISO~\cite{CAISO3100} & UC with MOC & SG/SC & Uplift \\
NESO~\cite{NGESO2023Stability} & IAUC & SG/SC, IBR & Marginal pricing \\
EirGrid~\cite{Eirgrid} & IAUC & SG/SC & Contract pricing \\
AEMO~\cite{AEMO2024FCESS} & IAUC & SG/SC & Marginal pricing \\
\midrule
Doherty \textit{et al.}~\cite{doherty2005frequency} & IAED (Deterministic) & SG & Marginal pricing \\
Hu \textit{et al.}~\cite{hu2023inertia} & IAED (Stochastic) & SG, Wind turbine, Battery & Marginal pricing \\
Doherty \textit{et al.}~\cite{paturet2020stochastic} & IAUC (Stochastic) & SG, IBR & -- \\
Doherty \textit{et al.}~\cite{badesa2019simultaneous} & IAUC (Stochastic) & SG, Wind turbine & -- \\
Liang \textit{et al.}~\cite{liang2022inertia} & IAUC (Chance constraint) & SG, Wind turbine & Marginal pricing \\
Qiu \textit{et al.}~\cite{qiu2024market} & IAUC (Deterministic) & SG, IBR & Marginal pricing \\
Paturet \textit{et al.}~\cite{paturet2020economic} & IAUC (Deterministic) & SG, IBR & Uplift, Marginal pricing  \\
Lu \textit{et al.}~\cite{lu2024convex} & IAUC (Deterministic) & SG, IBR & Convex hull pricing \\
\midrule
\textbf{This work} & IAUC (Chance constraint) & SG, PV, Wind, Battery & Uplift, Marginal pricing, Approximate convex hull pricing, Average incremental pricing \\
\bottomrule
\multicolumn{4}{l}{\scriptsize * MOC: Minimum Online Commitment, IAUC: Inertia-Aware Unit Commitment, IAED: Inertia-Aware Economic Dispatch,}\\
\multicolumn{4}{l}{\scriptsize\hspace{0.22cm}SG: Synchronous Generator, SC: Synchronous Condenser, IBR: Inverter-Based Resource}
\end{tabular}
\vspace{-3mm}
\end{table*}

System operators (SOs) have traditionally relied on synchronous machines to supply rotational inertia for grid stability.
The Electric Reliability Council of Texas (ERCOT) mitigates inertia shortages by deploying reliability-must-run (RMR) units when system inertia falls below 105 GW·s \cite{ERCOT2021}. These high-cost units operate at minimum output to limit market effects and are compensated through cost-recovery contracts \cite{ERCOT2024}.
The California ISO (CAISO) commits minimum-online-commitment (MOC) units in real time to maintain inertia and prevent transient instability \cite{CAISO3100}; associated losses are recovered via uplift payments \cite{CAISO2012BCR}.
In 2023, EirGrid procured 10 000 MW·s of inertial support from synchronous condensers through long-term contracts \cite{Eirgrid} and plans to provide virtual inertia by 2030 \cite{eirgrid2022lcis}.

To address declining synchronous inertia, several SOs are adopting IBRs and developing market mechanisms to support their economic viability.
The National Energy System Operator (NESO) has established a stability market that procures system-stability services through long-, mid-, and short-term contracts \cite{NGESO2023Stability}.
NESO requires IBRs to respond to RoCoF of 2 Hz/s within 5 ms \cite{neso2021tpr}.
In Australia, the Wholesale Electricity Market co-optimizes energy and frequency control services in real time via the Fast Contingency Essential System Services \cite{WEMRules2025}.
Both synchronous machines and IBRs can provide RoCoF control, and IBRs must remain online during disturbances with RoCoF up to 2 Hz/s over 250 ms \cite{AEMO2024FCESS}.

Several studies have explored strategies to ensure adequate inertia and assess its economic value.
Doherty et al. \cite{doherty2005frequency} proposed an inertia market based on provider bids and developed an inertia-aware economic-dispatch model, though their formulation neglects the stochastic nature of RESs under high penetration.
To address this, several studies \cite{hu2023inertia, paturet2020stochastic, badesa2019simultaneous} introduce stochastic market-design methods incorporating inertia services and requirements, while Liang et al. \cite{liang2022inertia} present a chance-constrained unit-commitment (UC) model with inertia constraints.
However, these studies still overlook uncertainties arising from heterogeneous RESs.

While \cite{liang2022inertia, hu2023inertia, qiu2024market} incorporate inertia services into market clearing mechanisms, they have focused on methodologies for inertia pricing and have not provided in-depth analysis comparing other pricing methods, which are essential when determining remuneration methods for inertia provision. Moreover, when employing a UC-based model, the inertia price derived from the dual variables of the inertia requirement can be zero. Uplift is also considered as a means to ensure revenue adequacy for inertia providers~\cite{paturet2020economic}.
However, uplift have notable drawbacks, as they fail to provide effective market price signals for system planning and operation, thereby disturbing market equilibrium.
To address this issue, Lu et al. \cite{lu2024convex} proposed an inertia-pricing method based on convex-hull pricing, which relaxes the non-convexity of SG commitment variables to derive inertia prices. However, convex-hull pricing does not guarantee polynomial-time convergence, and non-smooth optimization techniques such as sub-gradient methods often terminate before reaching optimality \cite{hua2016convex}.
The average incremental pricing (AIP) method \cite{o2019essays} offers an alternative by converting non-load and start-up costs into average incremental costs based on UC results and relaxing commitment variables.
Table \ref{tab:inertia_comparison} summarizes existing inertia-aware operation strategies and remuneration methods, including those in practice, prior research, and this work.


In this paper, we enhance an inertia-aware UC model that co-optimizes energy, reserves, and inertia provision in a decarbonized power system. We further develop and compare remuneration methods for inertia provision to provide comprehensive insights into developing sustainable market mechanisms. The main contributions are summarized as follows:
\begin{enumerate} 
    \item \textit{Inertia-aware CC-UC:} 
        Building on~\cite{liang2022inertia}, we extend the CC-UC model with inertia by incorporating time-coupling constraints of synchronous generators (SGs). This enhancement ensures reliable operation under high renewable penetration with heterogeneous RESs.
    \item \textit{Implementation and Comparative Analysis of Various Remuneration Methods:}
    We implement and compare remuneration schemes for inertia provision—uplift, marginal pricing (MP), approximate convex hull pricing (aCHP), and average incremental pricing (AIP)—within the inertia-aware UC framework.
    Analysis results show that aCHP minimizes uplift and ensures revenue adequacy, whereas MP and AIP fail to do so without additional compensation, with MP also lacking a clear price signal for inertia provision.
    \item \textit{Practical Validation:} 
        Using the IEEE 118-bus and Korean 193-bus test systems, we evaluate the operational and economic impacts of inertia-aware market design. 
        Results confirm that inertia-aware operation improves frequency stability and provide practical guidelines for developing inertia remuneration frameworks suitable for future low-inertia systems.
\end{enumerate}

\section{Inertia-aware Unit Commitment}

\subsection{Modeling Uncertainty of Generation and Inertia}
The active power output of photovoltaic (PV) and wind turbine (WT) units is represented as a random variable composed of a forecast value and forecast error:
\begin{subequations}
    \begin{align}
        \bm{P}_{pi,t} &= P_{pi,t} - \bm{\omega}_{pi,t}, \\
        \bm{P}_{wi,t} &= P_{wi,t} - \bm{\omega}_{wi,t},
    \end{align}
\end{subequations}
where $\bm{\omega}_{pi,t}\!\sim\!\mathcal{N}(m_{pi,t},\sigma_{pi,t}^2)$ and $\bm{\omega}_{wi,t}\!\sim\!\mathcal{N}(m_{wi,t},\sigma_{wi,t}^2)$ denote normally distributed forecast errors.  
Assuming the forecast errors are mutually independent~\cite{kargarian2015chance}, the system-wide generation uncertainty is
\begin{align}
    \bm{\Omega}_{rt}=\sum\nolimits_{i\in\mathcal{I}}\!(\bm{\omega}_{pi,t}+\bm{\omega}_{wi,t}) \sim \mathcal{N}(\mathrm{M}_{rt},\Sigma_{rt}^2),
\end{align}
where $\mathrm{M}_{rt}\!=\!\sum_i(m_{pi,t}\!+\!m_{wi,t})$ and $\Sigma_{rt}^2\!=\!\sum_i(\sigma_{pi,t}^2\!+\!\sigma_{wi,t}^2)$.

Similarly, the inertia contributions of PV and WT are uncertain and expressed as
\begin{subequations}
    \begin{align}
        \bm{H}_{pi,t} &= H_{pi,t} - \bm{\omega}_{phi,t},\\
        \bm{H}_{wi,t} &= H_{wi,t} - \bm{\omega}_{whi,t},
    \end{align}
\end{subequations}
where $\bm{\omega}_{phi,t}\sim\mathcal{N}(m_{phi,t},\sigma_{phi,t}^2)$ and $\bm{\omega}_{whi,t}\sim\mathcal{N}(m_{whi,t},\sigma_{whi,t}^2)$.
The aggregate inertia uncertainty across the system is
\begin{align}
    \bm{\Omega}_{ht}=\sum\nolimits_{i\in\mathcal{I}}(\bm{\omega}_{phi,t}+\bm{\omega}_{whi,t}) \sim \mathcal{N}(\mathrm{M}_{ht},\Sigma_{ht}^2).
\end{align}
Because PV inertia is provided via deloading~\cite{dreidy2017inertia}, it is treated as independent of PV power output and wind inertia. The deterministic equivalent formulation requires only that these random variables follow normal distributions without assuming dependency~\cite{liang2022inertia}.

\subsection{Formulation of the Inertia-aware CC-UC}
The market model in~\eqref{eq:model_1} represents the inertia-aware CC-UC, including SG, energy storage (ES), PV, and WT.
\begin{subequations}\label{eq:model_1}
\allowdisplaybreaks
    \begin{align}
    \min_{\substack{\mathclap{\Xi}}}\quad&\sum\limits_{t \in \mathcal{T}} \sum\limits_{i \in \mathcal{I}} \mathbb{E}_{\bm{\Omega}_{rt}} \big[c_{gi}(\bm{P}_{gi,t}, u_{gi,t}, v_{gi,t})\big] \label{eq:model_1_a}\\
    \mathrm{s.t.}\quad&\forall{i}\in\mathcal{I}, \forall{j}\in\mathcal{N}_i\, \forall t\in\mathcal{T}: \nonumber \\
    & u_{gi,t},~v_{gi,t},~w_{gi,t}, \in \{ 0,1\},  \label{eq:model_1_b}\\
    & u_{gi,t}-u_{gi,t-1} = v_{gi,t}-w_{gi,t}, \quad \forall t \in [2, T],\!\!\! \label{eq:model_1_c}\\
    &\textstyle\sum_{\tau=t-TU_{gi}+1}^{t} \!\!v_{gi,\tau} \leq u_{gi,t}, \quad \forall t \in [TU_{gi}, T],\!\!\! \label{eq:model_1_d}\\
    &\textstyle\sum_{\tau=t-TD_{gi}+1}^{t}\!\! w_{gi,\tau} \leq 1 - u_{gi,t}, ~ \forall t \in [TD_{gi}, T],\!\!\! \label{eq:model_1_e}\\ 
    & -\!RD_{gi} \leq p_{gi,t} -\! p_{gi,t-1} \leq RU_{gi}: (\upsilon_{gi,t}^{-},\upsilon_{gi,t}^{+}), \label{eq:model_1_f}\\
    & 0 \le \alpha_{gi,t} \le u_{gi,t}:\quad (\rho_{gi,t}^{-},\rho_{gi,t}^{+}), \label{eq:model_1_g}\\
    & \mathbb{P}_{\bm{\Omega}_{rt}} \!\! \left[ {\bm{P}_{gi,t} \le u_{gi,t} P_{gi}^{\max }} \right] \ge 1 - \epsilon_{gi}, \label{eq:model_1_h}\\
    & \mathbb{P}_{\bm{\Omega}_{rt}} \!\! \left[ { \bm{P}_{gi,t} \ge u_{gi,t} P_{gi} ^{\min}} \right] \ge 1 - \epsilon_{gi}, \label{eq:model_1_i}\\
    & \mathbb{P}_{\bm{\Omega}_{rt}} \left[\bm{P}_{di,t}+ 2H_{ei,t} {f_{\max}^{\prime}} P_{ei}^{\max}/f_0 \le P_{ei}^{\max}  \right] \ge 1 - \epsilon_{di}, \label{eq:model_1_j}\\
    & \mathbb{P}_{\bm{\Omega}_{rt}} \left[\bm{P}_{ci,t}+ 2H_{ei,t}  {f_{\max }^{\prime}} P_{ei}^{\max}/f_0 \le P_{ci}^{\max}  \right] \ge 1 - \epsilon_{ci}, \label{eq:model_1_k}\\
    & e_{i,t} \le E_i^{\max} - 2H_{ei,t} \Delta f_{\max} P_{ei}^{\max}/f_0, \label{eq:model_1_l}\\
    & e_{i,t} \ge E_i^{\min } + 2H_{ei,t} \Delta f_{\max} P_{ei}^{\max}/f_0, \label{eq:model_1_m}\\
    & e_{i,t} =  e_{i,t-1} + \mathbb{E}_{\bm{\Omega}_{rt}} \left[\bm{P}_{ci,t} k_i -\bm{P}_{di,t}/k_i \right], \label{eq:model_1_n}\\
    & 0 \le \alpha_{ci,t},~\alpha _{di,t} \le 1, \label{eq:model_1_o}\\
    & H_{ei,t} \le H_{ei}^{\max}, \label{eq:model_1_p}\\
    & -F_{i,j}^{\mathrm{max}} \le B_{i,j} (\theta_{i,t}-\theta_{j,t}) \le F_{i,j}^{\mathrm{max}}, \label{eq:model_1_q}\\
    \begin{split}
    & P_{gi,t} + P_{di,t} - P_{ci,t} + P_{wi,t} + P_{pi,t} - D_{i,t} \\
    & = \textstyle\sum\nolimits_{j \in \mathcal{N}_i } B_{i,j} (\theta_{i,t} {\rm{-}} \theta_{j,t}):\quad (\lambda_{i,t}), \label{eq:model_1_r}
    \end{split}\\
    & \textstyle\sum\nolimits_{i \in \mathcal{I}} {\left( {\alpha_{gi,t} + \alpha_{di,t} - \alpha_{ci,t}} \right)}  = 1:\quad (\gamma_t),  \label{eq:model_1_s}\\
    \begin{split}
    & \mathbb{P}_{\bm{\Omega}_{ht}} \Big[ \textstyle\sum\limits_{i \in \mathcal{I}} \left( u_{gi,t} H_{gi} P_{gi}^{\max} + H_{ei,t} P_{ei}^{\max} \right. \\
    & \left. \!+ \bm{H}_{pi,t}\!P_{pi}^{\max}\!+\bm{H}_{wi,t}\!P_{wi}^{\max}\!\right)\!\ge\!P_{sys} H_{\min} \Big]\!\ge\!1\!-\!\epsilon_{hi},\hspace{-10mm} \label{eq:model_1_t}
    \end{split}\end{align}\end{subequations} where $\mathcal{I}$ is the set of nodes, $\mathcal{N}_i$ is the set of nodes that are connected to node $i$, $\mathcal{T}$ is set of time, 
$\;\; \mathclap{\Xi}=$ $\{p_{gi,t}, u_{gi,t}, v_{gi,t}, w_{gi,t}, \\ \alpha_{gi,t}, p_{di,t}, p_{ci,t}, \alpha_{di,t}, \alpha_{ci,t}, H_{ei,t}\}$ being the set of optimization variables.
Objective~\eqref{eq:model_1_a} minimizes the total expected generation cost $c_{gi}(\cdot)$. 
Constraints~\eqref{eq:model_1_b}–\eqref{eq:model_1_e} enforce minimum up/down times, with $u_{gi,t}$, $v_{gi,t}$, and $w_{gi,t}$ denoting the on/off, startup, and shutdown status of SGs, and $TU_{gi}$, $TD_{gi}$ the respective limits.
Ramping limits appear in~\eqref{eq:model_1_f}.
The reserve participation factor $\alpha_{gi,t}$ is bounded in~\eqref{eq:model_1_g}.
Chance constraints~\eqref{eq:model_1_h}–\eqref{eq:model_1_i} ensure SG outputs $\bm{P}_{gi,t}$ remain within $[P_{gi}^{\min},P_{gi}^{\max}]$.
Equations~\eqref{eq:model_1_j}–\eqref{eq:model_1_k} limit ES charging/discharging to provide sufficient inertial response under the maximum RoCoF $f_{\max}^{\prime}$, while~\eqref{eq:model_1_l}–\eqref{eq:model_1_m} maintain safety margins for the frequency nadir $\Delta f_{\max}$.
Equation~\eqref{eq:model_1_n} updates ES energy considering expected charge/discharge and forecast errors.
Constraint~\eqref{eq:model_1_o} bounds ES reserve participation ($0$–$1$), and~\eqref{eq:model_1_p} defines ES inertia provision.
Constraints~\eqref{eq:model_1_q}-\eqref{eq:model_1_r} are DC power flow constraints for the line thermal limits, which yields locational marginal prices when line congestion occurs. 
Constraints~\eqref{eq:model_1_q}–\eqref{eq:model_1_r} represent DC power flow and thermal limits, yielding LMPs under congestion.
Constraint~\eqref{eq:model_1_s} ensures total reserve participation, and~\eqref{eq:model_1_t} enforces inertia adequacy from ESs, and RESs considering forecast uncertainty.

\subsection{Deterministic Reformulation of the Inertia-aware CC-UC}
The expectation ($\mathbb{E}$) and probability ($\mathbb{P}$) terms in~\eqref{eq:model_1} are reformulated into tractable deterministic expressions.

\subsubsection{Expected Generation Cost}
The generation cost of an SG is approximated quadratically as
\[
c_{gi}(\bm{P}_{gi,t},\!u_{gi,t},\!v_{gi,t})
= a_{gi}\bm{P}_{gi,t}^2 \!+\! b_{gi}\bm{P}_{gi,t}
\!+\! u_{gi,t}c_{gi} \!+\! v_{gi,t}s_{gi},
\]
where $a_{gi}$, $b_{gi}$, and $c_{gi}$ are cost coefficients and $s_{gi}$ is the start-up cost.  
Using $\mathbb{E}[\bm{x}^2]=\mathbb{E}[\bm{x}]^2+\mathrm{Var}[\bm{x}]$, the expected generation cost becomes
\begin{align}
\begin{split}
C_{gi,t} =& a_{gi}\!\left[(P_{gi,t}\!+\!\mathrm{M}_{pt}\alpha_{gi,t})^2
+ \Sigma_{pt}^2\alpha_{gi,t}^2\right]
\\ 
&+ b_{gi}(P_{gi,t}\!+\!\mathrm{M}_{pt}\alpha_{gi,t})
+ u_{gi,t}c_{gi}+v_{gi,t}s_{gi}. 
\end{split}\label{eq:cstftngi}
\end{align}

\subsubsection{Deterministic Chance Constraints}
Assuming normally distributed random variables, the inverse CDF $\Phi^{-1}(\cdot)$ converts probabilistic limits to deterministic ones~\cite{dvorkin2019chance}.  
Define auxiliary parameters:
\[
\hat\delta_{gi}\!=\!\Phi^{-1}(1-\epsilon_{gi})\Sigma_{rt}-\mathrm{M}_{rt},~~
\hat\delta_{di}\!=\!\Phi^{-1}(1-\epsilon_{di})\Sigma_{rt}-\mathrm{M}_{rt},
\]
\[
\hat\delta_{ci}\!=\!\Phi^{-1}(1-\epsilon_{ci})\Sigma_{rt}-\mathrm{M}_{rt},~~
\hat\delta_{hi}\!=\!\Phi^{-1}(1-\epsilon_{hi})\Sigma_{ht}-\mathrm{M}_{ht}.
\]
Substituting these parameters transforms \eqref{eq:model_1_h}–\eqref{eq:model_1_k}, \eqref{eq:model_1_n}, and \eqref{eq:model_1_t} into their deterministic counterparts. 

Then all probabilistic constraints in~\eqref{eq:model_1} are replaced with tractable equivalents:
\begin{subequations}
\begin{align}
\min\;&\sum_{t\in\mathcal{T}}\sum_{i\in\mathcal{I}}C_{gi,t} \label{model_determined_a}\\
\mathrm{s.t.}\;&\text{\eqref{eq:model_1_b}–\eqref{eq:model_1_g}, \eqref{eq:model_1_l}–\eqref{eq:model_1_m}}, \nonumber\\
& P_{gi,t}\le u_{gi,t}P_{gi}^{\max}-\hat\delta_{gi}\alpha_{gi,t}:\quad (\mu^+_{i,t}), \label{eq:deterministicModel_b}\\
& P_{gi,t}\ge u_{gi,t}P_{gi}^{\min}+\hat\delta_{gi}\alpha_{gi,t}:\quad (\mu^-_{i,t}), \label{eq:deterministicModel_c}\\
& P_{di,t}+2H_{ei,t}f_{\max}^{\prime}P_{ei}^{\max}/f_0
\le P_{ei}^{\max}-\hat\delta_{di}\alpha_{di,t}, \label{eq:deterministicModel_d}\\
& P_{ci,t}+2H_{ei,t}f_{\max}^{\prime}P_{ei}^{\max}/f_0
\le P_{ci}^{\max}-\hat\delta_{ci}\alpha_{ci,t}, \label{eq:deterministicModel_e}\\
\begin{split}
    & e_{i,t}=e_{i,t-1}+(P_{ci,t}+\mathrm{M}_{pt}\alpha_{ci,t})k_i
    \\&\hspace{4mm}-(P_{di,t}+\mathrm{M}_{pt}\alpha_{di,t})/k_i, \label{eq:deterministicModel_f}
\end{split}\\
\begin{split}
 &\small\sum_{i\in\mathcal{I}}\!\Big(u_{gi,t}H_{gi}P_{gi}^{\max}
\!+\!H_{ei,t}P_{ei}^{\max}
\!+\!(H_{pi,t}\!+\!\hat\delta_{hi})P_{pi}^{\max}\hspace{-8mm}
\\&\hspace{4mm}+(H_{wi,t}+\hat\delta_{hi})P_{wi}^{\max}\Big)
\!\ge\!P_{sys}H_{\min}:\quad (\chi_t). \label{eq:deterministicModel_g}
\end{split}
\end{align}\label{eq:deterministicModel}
\end{subequations}

\section{Remuneration methods for Inertia Provision}

\subsection{Uplift}
Uplift serves as a compensation mechanism for units that incur additional costs to maintain system reliability or provide inertia. 
For RMR units, uplift reimburses the generation cost that is not recovered through market settlements.
Similarly, IBRs that provide synthetic inertia via deloading experience opportunity costs corresponding to the curtailed active power. 
The uplift for RMR units and IBRs is formulated as follows:
\begin{subequations}
    \begin{align}
    \Psi_{gi} &= \textstyle\sum\nolimits_{t \in \mathcal{T}}\big({c}_{gi}(p_{gi,t})-\Lambda_t p_{gi,t}\big),
    \label{eq:cost_compensation} \\
    \Pi_{gi} &= \textstyle\sum\nolimits_{t \in \mathcal{T}}\Lambda_t (p_{gi,t}^{\mathrm{mppt}} - p_{gi,t}), \label{eq:opportunity_cost}
    \end{align}
\end{subequations}
where $\Psi_{gi}$ represents the uplift payment to unit $i$ for unrecovered production costs, 
$\Pi_{gi}$ denotes the opportunity cost for an IBR providing deloaded synthetic inertia, 
$\Lambda_t$ is the market clearing price at time $t$, $p_{gi,t}$ is the actual power output, and $p_{gi,t}^{\mathrm{mppt}}$ is the maximum available power.

\subsection{Marginal pricing}
The deterministic formulation in~\eqref{eq:deterministicModel} is a mixed-integer quadratic program (MIQP).
After obtaining the optimal commitment decisions $u_{gi,t}^{*}$, the problem can be reformulated as an equivalent convex quadratic program (QP) by fixing the commitment variables as follows:
\begin{subequations}\label{eq:convexModel}
    \begin{align}
    \min \quad&\textstyle\sum\nolimits_{t \in \mathcal{T}} \textstyle\sum\nolimits_{i \in \mathcal{I}}  C_{gi,t} \label{model_convexModel_a}\\
    \mathrm{s.t. }\quad& 
    \text{\eqref{eq:model_1_b}\textendash\eqref{eq:model_1_g}, \eqref{eq:model_1_l}\textendash\eqref{eq:model_1_m}, \eqref{eq:deterministicModel_b}\textendash\eqref{eq:deterministicModel_g}}, \\
    & u_{gi,t}=u_{gi,t}^*. \label{model_convexModel_b}
    \end{align}
\end{subequations}
We can derive marginal prices from dual variables of \eqref{eq:convexModel}, which are the market prices \cite{liang2022inertia}. 
\subsubsection{Energy price in MP}
Energy price, represented by dual variable $\lambda_{i,t}$ associated with constraint~\eqref{eq:model_1_r}, is derived from the Karush–Kuhn–Tucker (KKT) optimality conditions as:
\begin{align}
\begin{split}
     \lambda_{i,t} =2a_{gi}(P_{gi,t} + ~&\mathrm{M}_{pt} \alpha_{gi,t}) + b_{gi}\\
    &+ \mu_{i,t}^{+} - \mu_{i,t}^{-} +\upsilon_{gi,t}^{+} - \upsilon_{gi,t}^{-},
    \end{split}\label{eq:lambda_i_t}
\end{align}
where $\mu_{i,t}^{+}$ and $\mu_{i,t}^{-}$ are the dual variables associated with lower/upper power bounds in \eqref{eq:deterministicModel_b}\textendash\eqref{eq:deterministicModel_e}, and $\upsilon_{gi,t}^{+}$ and $\upsilon_{gi,t}^{-}$ correspond to ramping limits in \eqref{eq:model_1_f}.
Marginal energy price $\lambda_{i,t}$ depends on the generator cost coefficients $a_{gi}$ and $b_{gi}$, as defined in~\eqref{eq:cstftngi}, and reflects both the quadratic production cost and the binding limits of generation and reserve capacities.

\subsubsection{Reserve price in MP}
Similarly, the reserve price, represented by dual variable $\gamma_t$ associated with reserve participation constraint in \eqref{eq:model_1_s}, is expressed as:
\begin{align}
    \gamma_t & = \frac{{\textstyle\sum\limits_{i \in \mathcal{I}} { [ {b_{gi}\mathrm{M}_{pt} + ({\mu_{i,t}^{+}  + \mu _{i,t}^{-} } ){\hat \delta }_{gi} + \rho_{gi,t}^{+} - \rho_{gi,t}^{-}} ]/2a_{gi}}}}{{\sum\nolimits_{i \in \mathcal{I}} {1/(2a_{gi})} }} \nonumber\\
    & + \frac{\mathrm{M}_{pt} P_{gi,t} + { \left [1 - \sum\nolimits_{i \in \mathcal{I}} {( {{\alpha _{di,t}} - {\alpha_{ci,t}}} ) }\right] \left( {\Sigma_{pt} ^2 + \mathrm{M}_{pt} ^2} \right) }}{{\sum\nolimits_{i \in \mathcal{I}} {1/\left( {2a_{gi}} \right)} }},
    \label{eq:gamma_t}
\end{align}
where $\rho_{gi,t}^{+}$ and $\rho_{gi,t}^{-}$ denote the dual variables corresponding to \eqref{eq:model_1_g}.
Similar to the energy price, reserve price $\gamma_t$ is influenced by generator cost coefficients $a_{gi}$ and $b_{gi}$, reflecting the marginal cost of providing reserve capacity.

\subsubsection{Inertia price in MP}
Based on dual variable $\chi_t$ associated with constraint~\eqref{eq:model_1_t}, the inertia price is formulated as:
\begin{align}
\begin{split}
    &\chi_t = \\
    &\frac{
        {\sum\nolimits_{i \in \mathcal{I}} {\left[{u_{gi,t}}({c_{gi}} - \mu_{i,t}^{+} + \mu_{i,t}^{-} + \kappa_{i,t} - \rho_{i,t}^+)\right]} }
    }{
        \scriptstyle{P_{\!sys} H_{\!\min} \!-\! \sum\limits_{i \in \mathcal{I}} {[ {H_{ei,t}P_{ei}^{\max} \!+\! (H_{pi,t} + {{\hat \delta}_{pi}})P_{pi}^{\max} \!+\! (H_{wi,t} + {{\hat \delta}_{wi}})P_{wi}^{\max}} ]}}
    },\hspace{-2mm}
\end{split}\label{eq:chi_t}
\end{align}
where $\kappa_{i,t}$ denotes the dual variable corresponding to \eqref{model_convexModel_b}.
When constraint~\eqref{eq:deterministicModel_g} is binding, inertia price $\chi_t$ takes a non-zero value. 
It depends primarily on non-load cost $c_{gi}$ and dual variable $\kappa_{i,t}$ of the committed SG, and increases as the inertia contribution from generators decreases. 
Notably, start-up costs do not affect the inertia price.

\subsection{Approximated convex hull pricing (aCHP)}
The aCHP is a post-processing approach that determines prices after the unit commitment schedule is fixed, independently of real-time operation decisions~\cite{wang2016commitment}.
\subsubsection{Convexified single-hour cost function} 
To incorporate start-up costs into the price formation, the generation cost function in~\eqref{eq:cstftngi} is reformulated with the distributed start-up cost $\tilde{s}_{gi,t}$ as:
\begin{align}
\begin{split}
    \Tilde{C}_{gi,t}  =~& a_{gi} \left[ (P_{gi,t} + \mathrm{M}_{pt} \alpha _{gi,t})^2 + \Sigma_{pt}^2 \alpha_{gi,t}^2 \right] \\  
    &+ b_{gi}\left( {P_{gi,t}+\mathrm{M}_{pt} \alpha _{gi,t}} \right) + u_{gi,t}(c_{gi} + \Tilde{s}_{gi,t}),
\end{split}\label{eq:new_cost_function}
\end{align}
where the start-up cost $s_{gi}$ is allocated to online time $t$ for the committed generators, which are decided in~\eqref{eq:deterministicModel}, i.e., the distributed start-up cost $\Tilde{s}_{gi,t}$ is represented as:
\begin{align}
    s_{gi} = \textstyle\sum_t \Tilde{s}_{gi,t}, \quad t \in [t_{on}, t_{off}).
    \label{eq:allocated_startup_cost}
\end{align}
The reformulated generation cost $\Tilde{C}_{gi,t}$ is convexified by relaxing a commitment variable satisfying~\eqref{eq:deterministicModel_b},~\eqref{eq:deterministicModel_c} and:
\allowdisplaybreaks
\begin{align}
    0 \leq u_{gi,t} \leq 1, \quad \forall t\in\mathcal{T}, \forall i\in\mathcal{I}.
    \label{eq:relaxtion_commitment}
\end{align}
Different allocation rules for distributing $s_{gi}$ across online hours lead to varying price outcomes.

\subsubsection{Time decoupled convex equivalent}
The convex hull-based price at time $t$ is obtained by minimizing the sum of the convexified single-hour cost of generator $i$ as:
\allowdisplaybreaks
\begin{subequations}\begin{align}
    \min_{\substack{\mathclap{\Xi}}} \quad&\textstyle\sum\limits_{i \in \mathcal{I}} \Tilde{C}_{gi,t}  \\
     \mathrm{s.t.} \quad & \eqref{eq:model_1_f}, \eqref{eq:model_1_g}, \eqref{eq:model_1_q}-\eqref{eq:model_1_t},  \eqref{eq:deterministicModel_b}, \eqref{eq:deterministicModel_c}, \eqref{eq:deterministicModel_g}, \eqref{eq:relaxtion_commitment}, 
\end{align}\label{eq:def_aelmp}\end{subequations}
where $\;\mathclap{\Xi} \coloneqq \{p_{gi,t}, u_{gi,t}, \alpha_{gi,t}\}$ denotes the set of optimization variables.
The problem in~\eqref{eq:def_aelmp} is solved independently for each time interval.
For ramping constraints~\eqref{eq:model_1_f}, the previous generation level $p_{gi,t-1}$ is fixed from the solution of the preceding interval.
In addition, the charging/discharging power, energy level, and provided inertia of the energy storage (ES) unit are fixed based on the dispatch results from \eqref{eq:deterministicModel}, ensuring energy balance throughout the entire ES operational period.

\subsubsection{Energy price in aCHP}
By differentiating the Lagrangian dual of~\eqref{eq:def_aelmp} with respect to $P_{gi,t}$, energy price $\lambda_{i,t}$ is obtained as:
\begin{align}
\begin{split}
    \lambda_{i,t} = 2a_{gi}(P_{gi,t} &+ \mathrm{M}_{pt} \alpha _{gi,t}) + b_{gi} 
    \\& +\mu_{gi,t}^{+} - \mu_{gi,t}^{-} + \upsilon_{gi,t}^{+} - \upsilon_{gi,t}^{-}.
\end{split}\label{eq:conv_ep}
\end{align}
Equation~\eqref{eq:conv_ep} is identical to~\eqref{eq:lambda_i_t} in MP, showing that the quadratic cost coefficients $a_{gi}$ and $b_{gi}$ determine the energy price in both MP and aCHP.
Non-load and start-up costs do not affect the energy price.

\subsubsection{Reserve price in aCHP}
The reserve price $\gamma_t$ is derived by differentiating the Lagrangian dual of~\eqref{eq:def_aelmp} with respect to $\alpha_{gi,t}$:
\begin{align}
\begin{split}
    \gamma_t = ~& 2a_{gi} \left[ {\mathrm{M}_{pt} P_{gi,t}+\alpha_{gi,t} \left( {\Sigma_{pt}^2+ \mathrm{M}_{pt}^2} \right)} \right] +b_{gi} \mathrm{M}_{pt}  \\
    & + \mu_{i,t}^{+} {\hat \delta }_{gi} + \mu_{gi,t}^{-} {\hat \delta }_{gi} + \rho_{gi,t}^{+} - \rho_{gi,t}^{-}.
\end{split}\label{eq:conv_rp}
\end{align}
Substituting\eqref{eq:conv_rp} into the system-wide reserve balance constraint~\eqref{eq:model_1_s} yields the reserve price $\gamma_t$ in the same form as~\eqref{eq:gamma_t}.
Thus, in both aCHP and MP, reserve prices are unaffected by the non-load and start-up costs of generators.

\subsubsection{Inertia price in aCHP}
In aCHP, the binary commitment variables are continuously relaxed, which enables the convex reformulation of the problem and ensures that constraint~\eqref{eq:model_1_t} remains active under bounded conditions.
The inertia price $\chi_t$ is obtained by differentiating the Lagrangian dual of~\eqref{eq:def_aelmp} with respect to $u_{gi,t}$ as:
\begin{align}
     \chi_t = \frac{c_{gi} + \Tilde{s}_{gi,t} + \kappa_{i,t}^{+} - \kappa_{i,t}^{-} - \rho_{gi,t}^{+}  - \mu_{gi,t}^{+}P_{gi}^{\max} + \mu_{gi,t}^{-}P_{gi}^{\min}}{H_{gi} P_{gi}^{\max}}.
         \label{eq:deri_u}
\end{align}
Substituting~\eqref{eq:deri_u} into~\eqref{eq:deterministicModel_g} as equality yields:
\begin{align}\label{eq:conv_chi_t}
\begin{split}
    &\chi_t =\!\\ 
    &\! \frac{{
    \scriptstyle\sum\nolimits_{i \in \mathcal{I}} {\left[{u_{gi,t}}(c_{gi} + \Tilde{s}_{gi,t} - \mu_{i,t}^{+} + \mu_{i,t}^{-} + \kappa^{+}_{i,t} - \kappa^{-}_{i,t} - \rho_{i,t}^+)\right]} }
    }{
    {\scriptstyle P_{\!sys} H_{\!\min} \!-\! \scriptstyle\sum\limits_{i \in \mathcal{I}} {[ {H_{\!ei,t}P_{\!ei}^{\max} + (H_{wi,t} + {{\hat \delta}_{\!wi}}\!)P_{\!wi}^{\!\max} + (H_{pi,t} + {{\hat \delta}_{pi}})P_{pi}^{\max}} ]}}
    }.
\end{split}
\end{align}
Compared with~\eqref{eq:chi_t}, the \textit{inertia price} in aCHP includes the distributed start-up cost $\tilde{s}_{gi,t}$.
This means that the relaxation of commitment variables in aCHP allows start-up costs to be incorporated into inertia remuneration—unlike MP, which excludes them from inertia pricing.

\subsection{Average incremental pricing}
We extend the average incremental pricing (AIP) framework~\cite{o2019essays} to derive \textit{inertia prices} within the proposed inertia-aware market design.

\subsubsection{Average incremental costs function} 
The average incremental cost function $\Hat{C}_{gi,t}$ incorporates both non-load and start-up costs into the generation cost of SGs defined as:
\begin{align}\label{eq:cost_function_aip}
\begin{split}
    \Hat{C}_{gi,t} =~& a_{gi} \left[ (P_{gi,t} + \mathrm{M}_{pt} \alpha _{gi,t})^2 + \Sigma_{pt}^2 \alpha_{gi,t}^2 \right] \\   
    & + \Hat{b}_{gi}\left( {P_{gi,t}+\mathrm{M}_{pt} \alpha _{gi,t}} \right),
\end{split}
\end{align}
where $\hat{b}_{gi}$ is the average cost coefficient. The non-load cost $c_{gi}$ and the start-up cost $s_{gi}$ are distributed across the total dispatched energy $\sum_{t\in[t_{\mathrm{on}},t_{\mathrm{off}})} P^{*}_{gi,t}$ as:
\begin{align}
    \Hat{b}_{gi} = b_{gi} + \frac{(\sum_{t\in [t_{\mathrm{on}}, t_{\mathrm{off}})}c_{gi}) + s_{gi}}{\sum_{t\in [t_{\mathrm{on}}, t_{\mathrm{off}})}P^{*}_{gi,t}}.
    \label{eq:aip_incremental_cost}
\end{align}
This formulation converts fixed and start-up costs into incremental costs, enabling a continuous cost representation.

\subsubsection{Time decoupled pricing run}
After fixing the SG commitment schedule obtained from~\eqref{eq:model_1}, prices are determined for each time interval by solving the following optimization:
\begin{subequations}\label{eq:def_aip}
\begin{align}
    \min_{\substack{\mathclap{\Xi}}} \quad &\sum\limits_{i \in \mathcal{I}} \Hat{C}_{gi,t}  \\
    \mathrm{s.t.} \quad & \eqref{eq:model_1_f}, \eqref{eq:model_1_g}, \eqref{eq:model_1_q}-\eqref{eq:model_1_t}, 
    \eqref{eq:deterministicModel_b}, \eqref{eq:deterministicModel_c}, \eqref{eq:deterministicModel_g}, \eqref{eq:relaxtion_commitment},
\end{align}
\end{subequations}
where $\mathclap{\Xi} = {p_{gi,t}, u_{gi,t}, \alpha_{gi,t}}$ represents the set of optimization variables.
Equation~\eqref{eq:def_aip} is solved independently for each hour with relaxed (continuous) commitment variables.
As a result, the generator operating at the minimum generation level determines the marginal price, while fixed costs are recovered through the incremental pricing structure.

\subsubsection{Energy price in AIP}
Differentiating the Lagrangian of \eqref{eq:def_aip} with respect to $P_{gi,t}$ yields:
\begin{align}
    \lambda_{i,t} \!=\! 2a_{gi}(P_{gi,t} \!+\! \mathrm{M}_{pt} \alpha_{gi,t}) \!+\! \Hat{b}_{gi} \!+\! \upsilon_{gi,t}^{+} \!-\! \upsilon_{gi,t}^{-} \!+\! \mu_{gi,t}^{+} \!-\! \mu_{gi,t}^{-}.
    \label{eq:aip_ep}
\end{align}
Relative to \eqref{eq:lambda_i_t} and \eqref{eq:conv_ep}, the only change is $b_{gi}\!\to\!\hat{b}_{gi}$, where $\hat{b}_{gi}$ embeds the average non-load and start-up costs of the marginal unit. 
Hence, AIP internalizes these fixed costs into the \textit{energy price} and is typically higher than MP and aCHP.

\subsubsection{Reserve price in AIP}
The \textit{reserve price} $\gamma_t$ is obtained by differentiating the Lagrangian of~\eqref{eq:def_aip} with respect to $\alpha_{gi,t}$:
\begin{align}
    \gamma_t= & 2a_{gi} \left[ {\mathrm{M}_{pt} P_{gi,t}+\alpha_{gi,t} \left( {\Sigma_{pt}^2+ \mathrm{M}_{pt}^2} \right)} \right] +\Hat{b}_{gi} \mathrm{M}_{pt}  \nonumber \\
    & + \mu_{i,t}^{+} {\hat \delta }_{gi} + \mu_{gi,t}^{-} {\hat \delta }_{gi} + \rho_{gi,t}^{+} - \rho_{gi,t}^{-}.
        \label{eq:aip_rp}
\end{align}
Compared with MP and aCHP, AIP introduces the average cost coefficient $\hat{b}_{gi}$, which embeds the non-load and start-up costs of the marginal generator.  
Therefore, the \textit{reserve price} in AIP reflects these fixed costs.

\subsubsection{Inertia price in AIP}
The \textit{inertia price} $\chi_t$ is obtained by differentiating the Lagrangian of~\eqref{eq:def_aip} with respect to $u_{gi,t}$ and substituting the result into~\eqref{eq:deterministicModel_g}:
\begin{align}\label{eq:aip_chi_t}
\begin{split}
    &\chi_t =\!\\
    &\frac{
    \scriptstyle{\sum\nolimits_{i \in \mathcal{I}} {\left[{u_{gi,t}}(- \mu_{i,t}^{+} + \mu_{i,t}^{-} + \kappa^{+}_{i,t} - \kappa^{-}_{i,t} - \rho_{i,t}^+)\right]} }
    }{
    \scriptstyle{P_{\!sys} H_{\min} - \sum\limits_{i \in \mathcal{I}} {[ {H_{\!ei,t}P_{\!ei}^{\max} + (\!H_{\!wi,t} + {{\hat \delta}_{\!wi}}\!)P_{\!wi}^{\max} + (\!H_{\!pi,t} + {{\hat \delta}_{\!pi}}\!)P_{\!pi}^{\max}} ]}}
    }.
\end{split}
\end{align}
Unlike aCHP, the \textit{inertia price} in AIP is unaffected by non-load and start-up costs, because these are already converted into the average incremental cost coefficient $\hat{b}_{gi}$.

\section{Case Studies}

\begin{figure}[t]
    \vspace{0mm}
    \centering    
    \includegraphics[width=1.0\columnwidth]{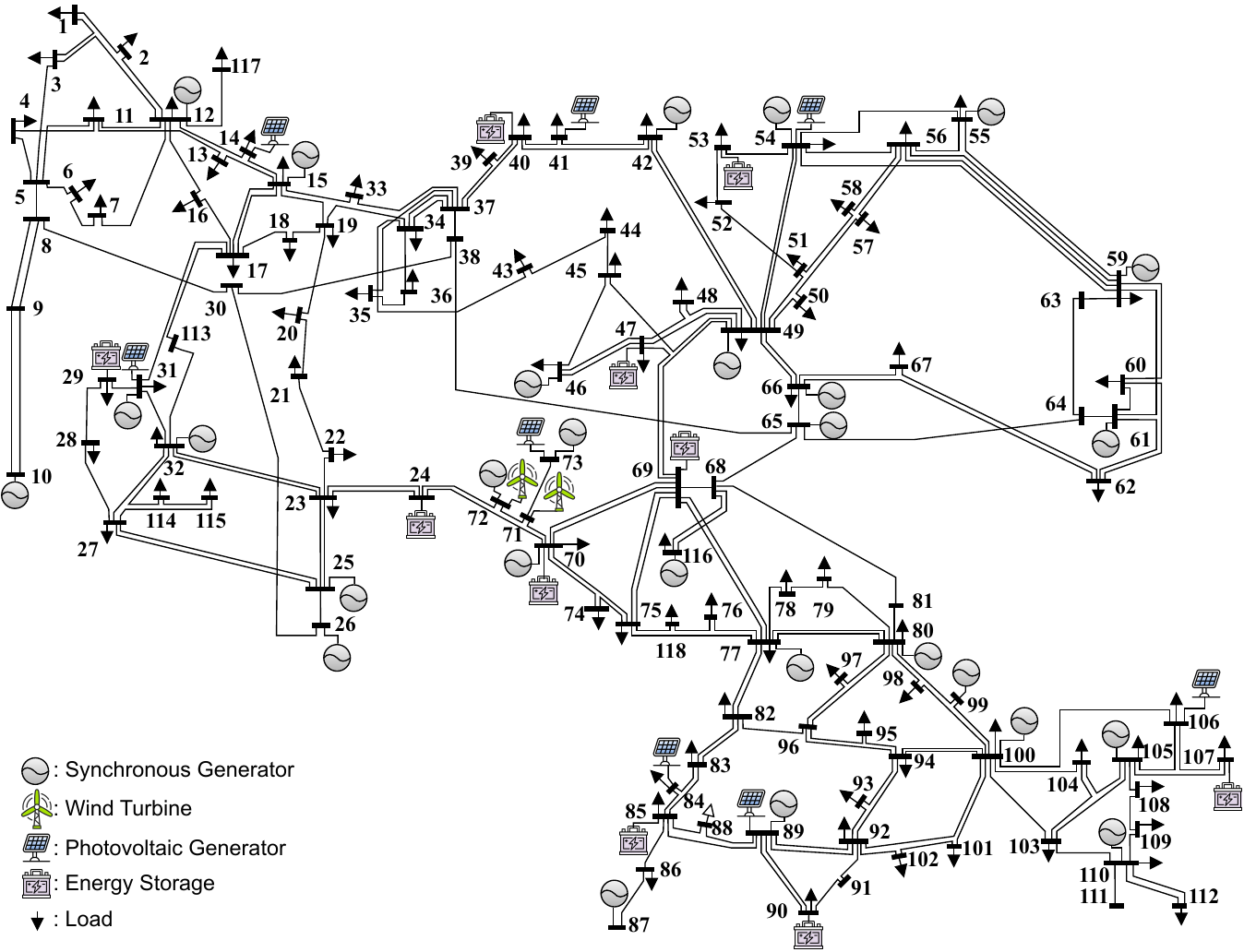}
    \vspace{-6mm}
    \caption{\small Illustration of the modified 118-bus network from~\cite{7904729}}
    \label{fig:network118}
    \vspace{-6mm}
\end{figure}

We employ the modified IEEE 118-bus test system~\cite{7904729} illustrated in Fig.\ref{fig:network118}, comprising 28 SGs, 8 PVs, 2 WTs, and 10 ESs providing energy, reserve, and inertia services. Ramping rates and minimum up/down times follow \cite{Carrion}. The inertia constants of SGs range from 3s to 10s depending on generator type~\cite{fernandez2019power}. 
RES generation profiles and load data are based on historical records from CAISO Fall 2023~\cite{Caiso}. 
The RES penetration levels are set as $\eta = \{20\%, 25\%, 30\%, 35\%, 40\%\}$. 
PV and WT units provide a uniform inertia constant of 3s by deloading 5\% and 10\% of their rated capacities, following \cite{villena2024assessment, dreidy2017inertia}.
ES units can provide inertia constants up to 11s. 
Table~\ref{T2:SystemParameters} summarizes the auxiliary parameters used in the simulations. 
All models are implemented in Julia using the JuMP package.
We set five scenarios for the numerical experiment:
\begin{itemize}
    \item \textbf{Base w/o inertia}: The inertia requirement constraint is not incorporated into the market model in~\eqref{eq:model_1}. After fixing the commitment variables, the price of energy and reserve are computed as dual variables.

    \item \textbf{Base w/ inertia (RMR)}: SGs not dispatched in the base w/o inertia case can be designated as RMR units to ensure adequate inertia levels. 
    These units are manually selected in ascending order of marginal cost until the inertia requirement is satisfied.  
    SGs with unrecovered costs are compensated through the uplift payment in~\eqref{eq:cost_compensation}, and inertia provided by IBRs is settled according to~\eqref{eq:opportunity_cost}.

    \item \textbf{MP}: The inertia-aware UC model in~\eqref{eq:model_1} is solved, and settlements are computed using prices for energy, reserve, and inertia obtained from~\eqref{eq:lambda_i_t},~\eqref{eq:gamma_t}, and~\eqref{eq:chi_t}.  
    Additional uplift payments are applied to ensure revenue adequacy.

    \item \textbf{aCHP}: The inertia-aware UC model in~\eqref{eq:model_1} is solved, and 
    settlements are determined using the prices derived from~\eqref{eq:conv_ep},~\eqref{eq:conv_rp}, and~\eqref{eq:conv_chi_t}. An uplift is also considered.

    \item \textbf{AIP}: The inertia-aware UC model in~\eqref{eq:model_1} is solved, and 
    settlements are based on the prices obtained from~\eqref{eq:aip_ep}–\eqref{eq:aip_chi_t}, including an uplift.
\end{itemize}

\begin{table}[b]\footnotesize
  \centering
    \captionsetup{justification=centering, labelsep=period, font=footnotesize, textfont=sc}
  \caption{Parameter setup for the case study}
    \begin{tabular}{p{6cm} p{1.5cm}}
    \toprule
    Parameter &
    Value\\
    \midrule
     Minimum equivalent inertia requirement, $H_{\min}$                            & 3.5s \\
     Maximum admissible RoCoF, $f_{\max}^{'}$                & 0.5Hz/s \\
     Maximum admissible frequency deviation, $\Delta f_{\max}$  & 0.55Hz \\
     Variance of the distribution, $\Sigma_{pt}$                                   & 1  \\
     Mean of the distribution, $\mathrm{M}$                                        & 0.5  \\
     Probability of resource's power limit violations, $\epsilon$                  & 0.05 \\
    \bottomrule
    \end{tabular}\vspace{-0mm}
  \label{T2:SystemParameters}
\end{table}

\begin{figure}[t]
    \centering
     \subfigure[$\eta = 20\%$]{%
        \includegraphics[width=0.48\linewidth]{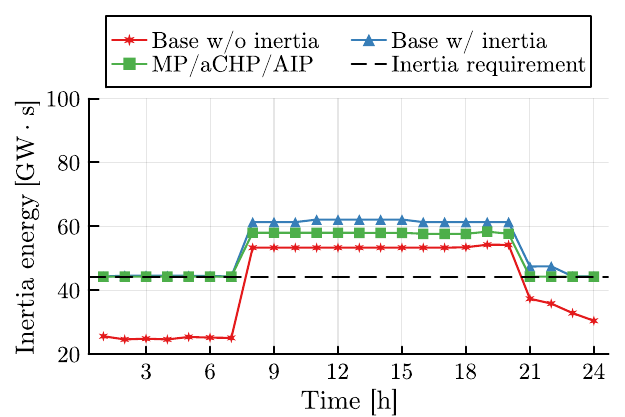}
        \label{fig:Sys_inertia_1}}
     \subfigure[$\eta = 40\%$]{
        \includegraphics[width=0.48\linewidth]{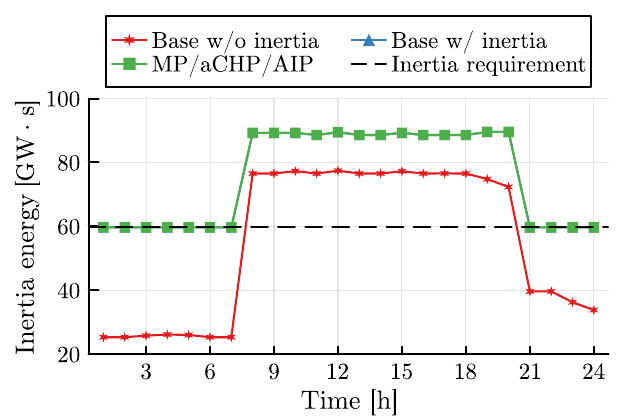}
        \label{fig:Sys_inertia_5}}
    \par\vspace{-3mm}
    \caption{\small
    System inertia levels provided by all generation resources (SGs and RES) under different scenarios and renewable penetration levels ($\eta$ = 20–40\%).
    The dashed line denotes the inertia requirement.
    }\label{fig:inertia_provision}
    \vspace{-6mm}
\end{figure}
\begin{figure}[t]%
    \centering
     \subfigure[At 01:00, $\eta = 20\%$]{%
        \includegraphics[width=0.48\linewidth]{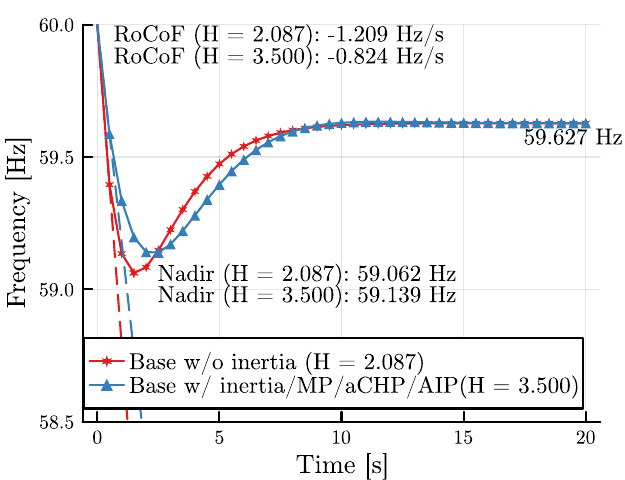}
        \label{fig:SFR_H_20_min}}\hfill
     \subfigure[At 01:00, $\eta = 40\%$]{
        \includegraphics[width=0.48\linewidth]{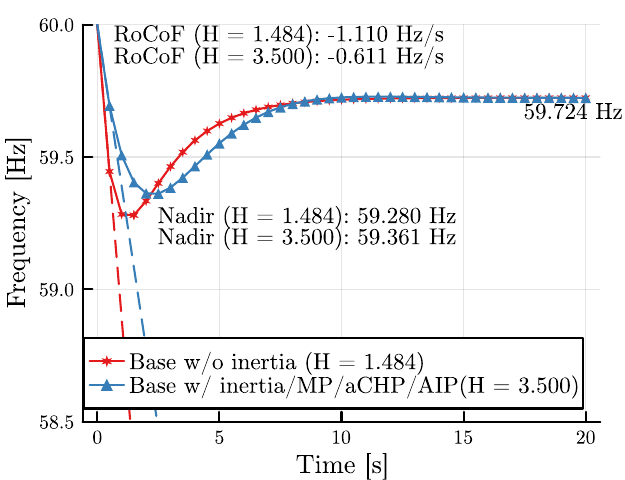}
        \label{fig:SFR_H_40_min}}\vfill
    \par\vspace{-3mm}
     \subfigure[At 08:00, $\eta = 20\%$]{
        \includegraphics[width=0.48\linewidth]{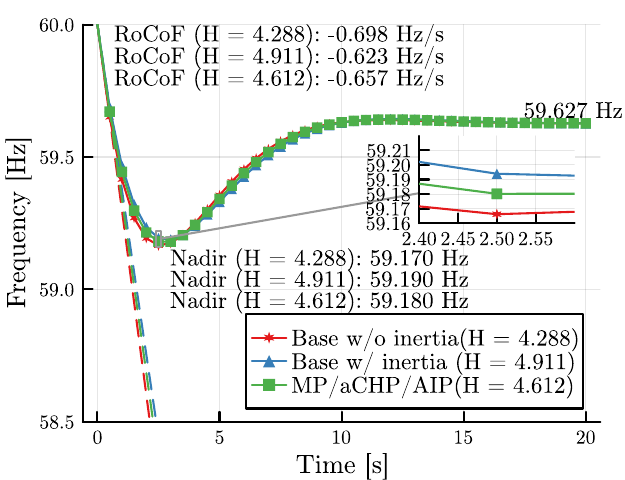}
        \label{fig:SFR_H_20_max}}\hfill
     \subfigure[At 08:00, $\eta = 40\%$]{
        \includegraphics[width=0.48\linewidth]{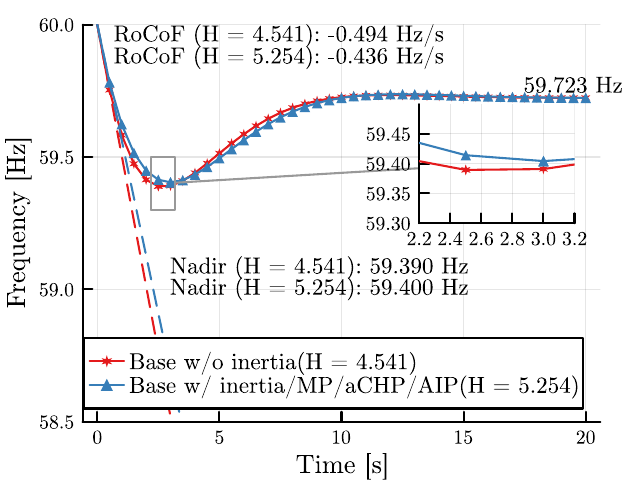}
        \label{fig:SFR_H_40_max}}\vfill
    \vspace{-3mm}
    \caption{\small 
    System frequency responses following a generation outage at different times (01:00 and 08:00) and renewable penetration levels ($\eta$ = 20–40\%).
    Each panel compares the Base case without inertia and cases with inertia provision under MP, aCHP, and AIP. RoCoF and frequency nadir values are annotated for each case.
    }
    \label{fig:SFR}
    \vspace{-6mm}
\end{figure}

\subsection{Effectiveness of Inertia-aware Operation}

We evaluate the effectiveness of inertia-aware operation (Base w/ inertia, MP, aCHP, and AIP) by comparing them against the Base w/o inertia case. 

\subsubsection{System inertia level}
Fig.~\ref{fig:inertia_provision} shows the system inertia energy levels for each case. 
Base w/o inertia case fails to meet the 44.28~GW$\cdot$s inertia requirement at $\eta = 20\%$ during off-peak hours (00:00-07:00 and 21:00-24:00), while all scenarios meet the requirement between 08:00 and 20:00. 
As $\eta$ increases, inertia requirements in Base w/o inertia are not met during off-peak periods, whereas all inertia-aware operation cases maintain adequate inertia throughout the day

\subsubsection{Frequency Stability}
The aggregate multi-machine system frequency response model in~\cite{anderson1990low} is used to analyze frequency dynamics following a generation outage. 
Using this model, we simulate the frequency response to the loss of the largest generator ($P^{\mathrm{max}} = 1{,}500$~MW) under each scenario.
Figs.~\ref{fig:SFR_H_20_min} and~\ref{fig:SFR_H_40_min} illustrate the frequency responses at two critical times: 01:00, when system inertia is at its minimum, and 08:00, when inertia is higher.  
Across all $\eta$, the frequency nadir improves compared with the Base w/o inertia case—from 59.062~Hz to 59.139~Hz and from 59.280~Hz to 59.361~Hz—representing a 0.13\% increase due to inertia-aware operation. 
RoCoF also improves by approximately 50\%, increasing from –1.209~Hz/s to –0.824~Hz/s and from –1.110~Hz/s to –0.611~Hz/s.  
At 08:00, however, the improvements in both nadir and RoCoF become negligible, as shown in Figs.~\ref{fig:SFR_H_20_max} and~\ref{fig:SFR_H_40_max}, because of enough system inertia.

\begin{figure}[t]
    \centering
     \subfigure[Base w/o inertia, $\eta = 20\%$]{%
        \includegraphics[width=0.48\linewidth]{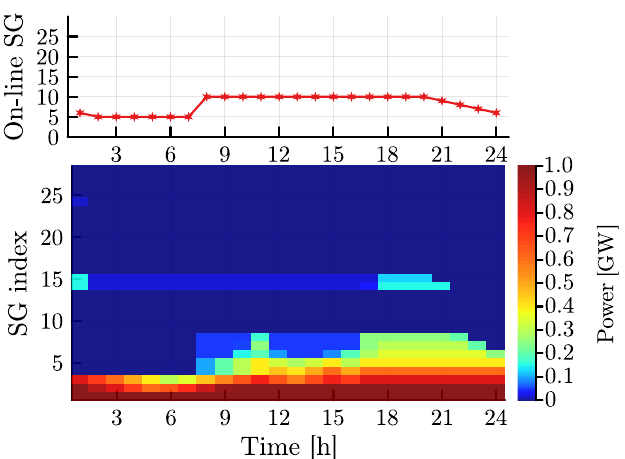}
        \label{fig:generation_base1}}\hfill
     \subfigure[Base w/o inertia, $\eta = 40\%$]{
        \includegraphics[width=0.48\linewidth]{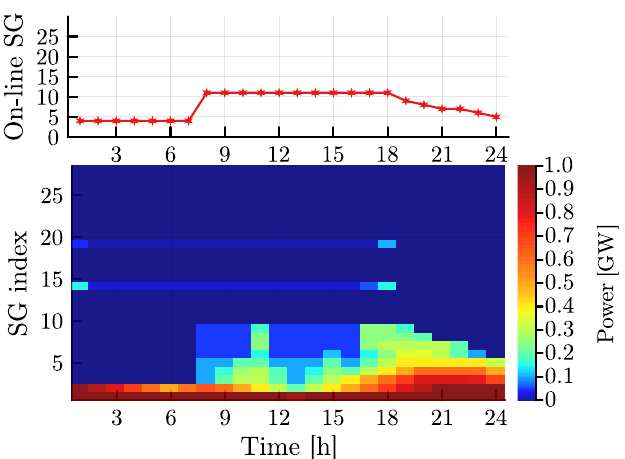}
        \label{fig:generation_base5}}\vfill
        \vspace{-3mm}
     \subfigure[Base w/ inertia, $\eta = 20\%$]{
        \includegraphics[width=0.48\linewidth]{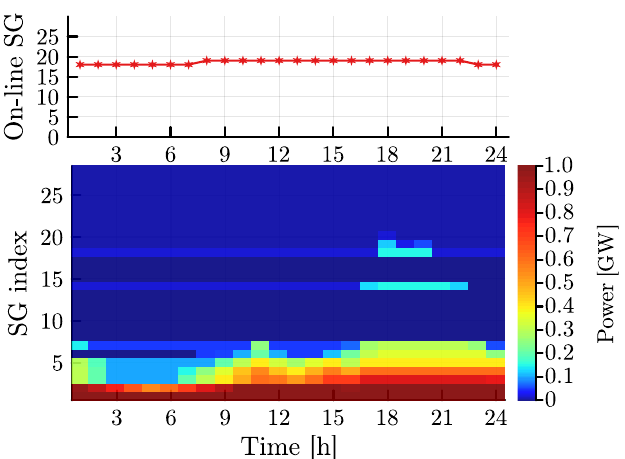}
        \label{fig:generation_uplift1}}\hfill
     \subfigure[Base w/ inertia, $\eta = 40\%$]{
        \includegraphics[width=0.48\linewidth]{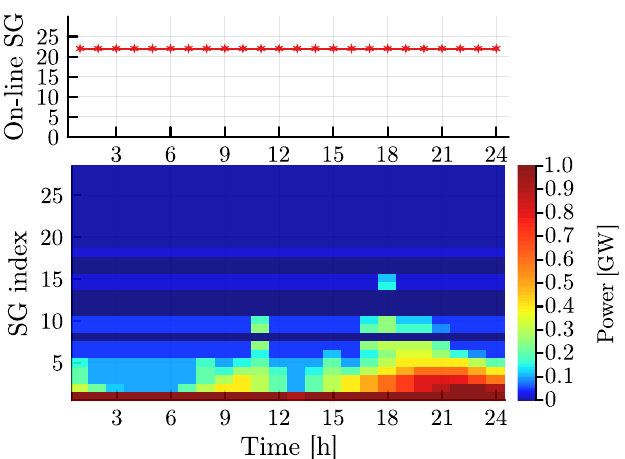}
        \label{fig:generation_uplift5}}\vfill
        \vspace{-3mm}
     \subfigure[MP/aCHP/AIP, $\eta = 20\%$]{
        \includegraphics[width=0.48\linewidth]{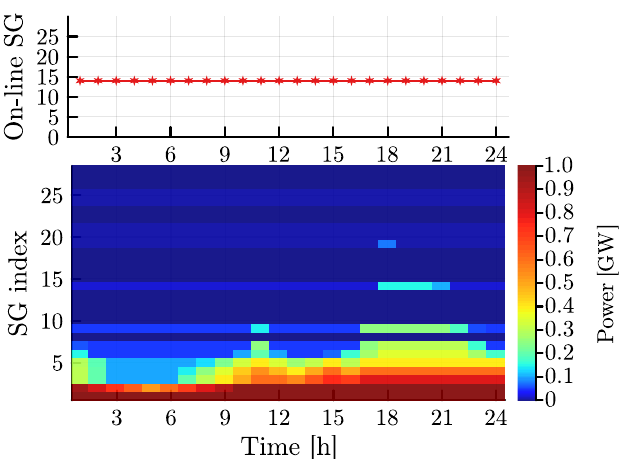}
        \label{fig:generation_pricing1}}\hfill
     \subfigure[MP/aCHP/AIP, $\eta = 40\%$]{
        \includegraphics[width=0.48\linewidth]{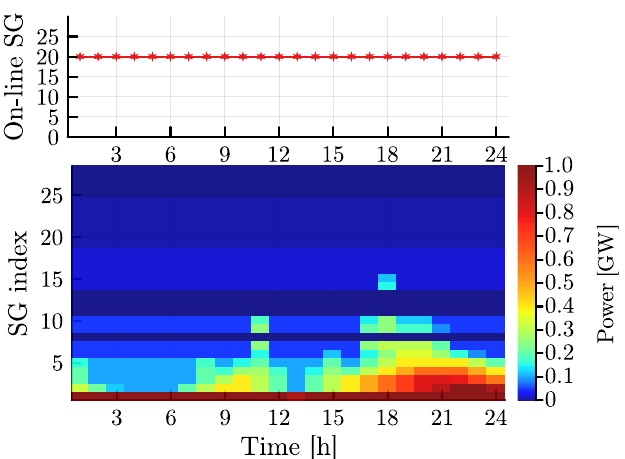}
        \label{fig:generation_pricing5}}\vfill
    \vspace{-3mm}
    \caption{\small
    Number of online SGs and their generation outputs under different scenarios and renewable penetration levels ($\eta$ = 20–40\%). 
    Each subplot shows the 24-hour variation of online SGs (top) and the corresponding dispatch levels (bottom).
    SG indices are sorted in ascending order of marginal cost.}
    \label{fig:generation}
    \vspace{-5mm}   
\end{figure}

\subsubsection{Commitment and generation of SG}
Figure~\ref{fig:generation} shows the number of online SGs and their generation outputs. 
In Base w/o inertia, the number of online SGs increases from 5 to 10 at 08:00 and from 4 to 11 at $\eta = 20\%$ and $\eta = 40\%$.
In Base w/ inertia, high-cost SGs (SG~16--26), which are not dispatched in Base w/o inertia, are committed as RMR units to meet the inertia requirement. 
Consequently, the number of online SGs is maintained at a minimum of 18 for $\eta = 20\%$ and 22 for $\eta = 40\%$. 
Because these RMR units operate at their minimum output levels, several mid-merit SGs (SG~6--15 at $\eta = 20\%$) are either not dispatched or operated at their minimum generation limits.
The total SG generation output decreases relative to the Base w/o inertia case.  
In MP, aCHP, and AIP, the number of online SGs is maintained at 14 for $\eta = 20\%$ and 20 for $\eta = 40\%$. 
These results demonstrate that inertia-aware operations can achieve the required inertia with fewer online SGs compared with Base w/ inertia, achieving more cost-effective scheduling. 
For example, SGs 19--21 and 24--26 are dispatched to provide inertia at $\eta = 20\%$, illustrating the efficiency of optimal inertia-aware unit commitment.

\begin{figure}[t]
    \centering
     \subfigure[Energy price in MP, aCHP, and AIP]{
        \includegraphics[width=1.0\linewidth]{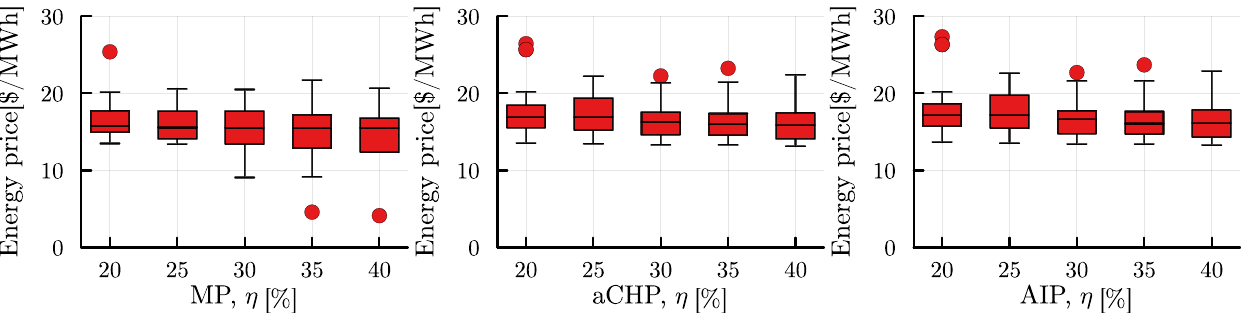}
        \label{fig:Energy_Price}}\vfill
        \vspace{-3mm}
     \subfigure[Reserve price in MP, aCHP, and AIP]{
        \includegraphics[width=1.0\linewidth]{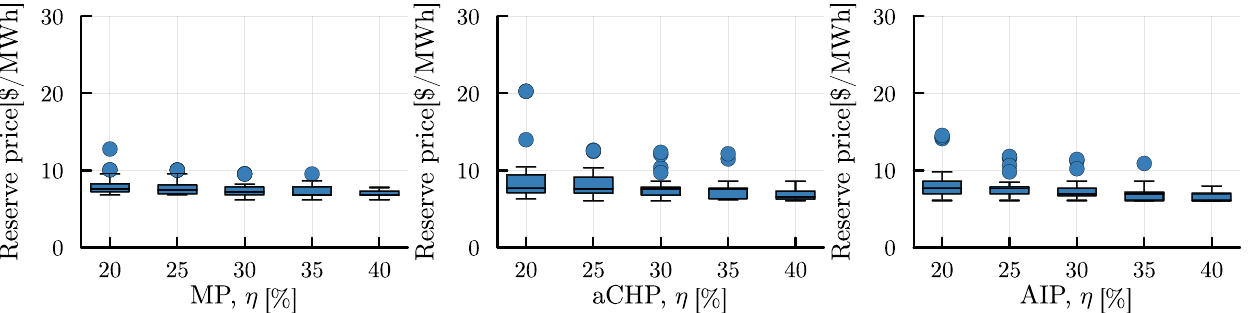}
        \label{fig:Reserve_Price}}\vfill
        \vspace{-3mm}
     \subfigure[Inertia prices in MP, aCHP, and AIP]{
        \includegraphics[width=1.0\linewidth]{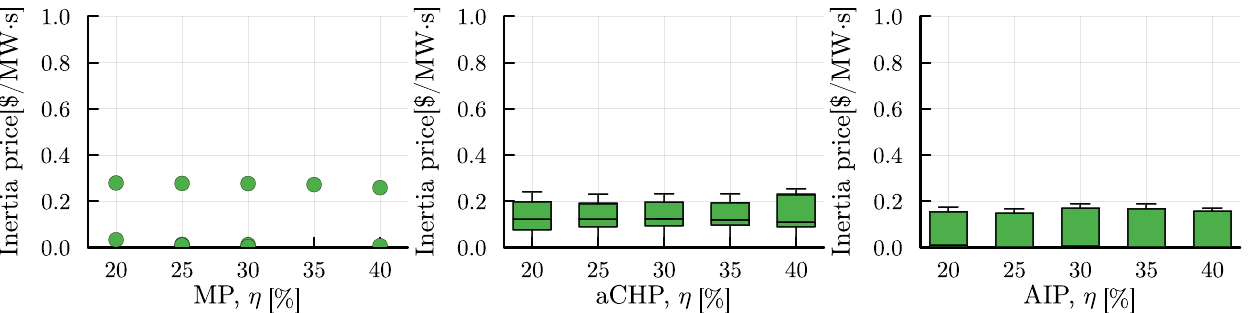}
        \label{fig:Inertia_Price}}\vfill    
    \vspace{-3mm}   
    \caption{\small Box plots of 24-hour energy, reserve, and inertia prices under MP, aCHP, and AIP schemes for different renewable penetration levels ($\eta$ = 20–40\%).
    Each box represents the 25th–75th percentile range with the median line inside.}
    \label{fig:prices}
    \vspace{-7mm}   
\end{figure}

\subsection{Comparison of Market Signals \& Remuneration Policies}

\subsubsection{Energy price}
Fig.~\ref{fig:Energy_Price} presents the energy prices under MP, aCHP, and AIP. 
Although the formulation of the energy price in MP and aCHP are identical, the average energy price in aCHP is 8.82\% higher than in MP. 
In aCHP, the minimum generation limit is relaxed to 0~MW, allowing the energy price to be determined at the minimum generation level of the marginal generator. 
For example, at $\eta = 20\%$ and 18:00, SG~20 operates under a 20~MW limit. In MP, it cannot set the price, yielding \$25.52/MWh—the marginal cost of SG~19—whereas in aCHP, SG~20 sets \$26.46/MWh at 2.37~MW.
In contrast, in aCHP, SG~20 determines the price at \$26.46/MWh at a dispatch level of 2.37~MW, below its minimum generation limit. 
The average energy price in AIP is 1.38\% higher than in aCHP because, as shown in~\eqref{eq:aip_ep}, AIP incorporates average non-load and start-up costs into the marginal price. 
For example, at $\eta = 20\%$, the marginal cost of SG~20 in AIP is \$26.65/MWh.
aCHP provides the most cost-reflective energy price because it captures the true marginal cost while preserving consistency with commitment decisions.  
AIP offers a slightly stronger long-term investment signal than MP by incorporating fixed costs, but it remains an approximation rather than a true marginal price formulation.

\subsubsection{Reserve price}
As shown in Fig.~\ref{fig:Reserve_Price}, reserve prices exhibit patterns similar to those of energy price across MP, aCHP, and AIP. 
With increasing $\eta$, average reserve prices decrease by about 2.5\%. 
High-cost SGs become constrained by minimum-generation limits and are excluded from determining the reserve price. 
Reserve prices in aCHP are higher than in MP because the marginal SG setting the reserve price in aCHP has a higher marginal cost. 
For example, at 18:00, the marginal SG in MP is SG~19, while in aCHP it is SG~20, leading to a price increase to \$13.99/MWh—\$1.29/MWh higher than in MP. 
If the marginal SG for energy and reserve are identical in both aCHP and AIP, the AIP price is higher because of embedding fixed costs. 
For example, at $\eta = 20\%$ and 02:00, when SG~3 provides all reserves, the reserve prices are \$7.62/MWh in aCHP and \$7.71/MWh in AIP.
Like energy price, AIP and aCHP yields a stronger long-term investment signal than MP by internalizing fixed costs within the pricing.

\subsubsection{Inertia price}
Fig.~\ref{fig:Inertia_Price} shows that inertia prices differ among the methods. 
In MP, inertia prices are almost zero for all $\eta$ because the inertia constraint in~\eqref{eq:deterministicModel_g} non-binding. 
In aCHP, relaxation of the commitment variables enables inertia to be treated as a continuous variable, making the inertia constraint binding and yielding non-zero inertia prices. 
AIP also produces non-zero inertia prices because commitment variables are relaxed during pricing. 
However, as AIP does not include non-load or start-up costs in inertia pricing, its inertia prices are consistently lower than those in aCHP.
From a market design perspective, aCHP provides the most efficient and transparent price signal for inertia provision because it fully reflects both operational and commitment-related costs within a convex framework. 
AIP, while simplifying cost representation through average incremental conversion, still offers a partial price signal that improves upon MP by internalizing fixed costs in energy and reserve pricing, though it underrepresents the true cost of inertia provision.

\begin{figure}[t]
    \centering
     \subfigure[Cost and revenue of SG 1 (base-load unit)]{%
        \includegraphics[width=1\linewidth]{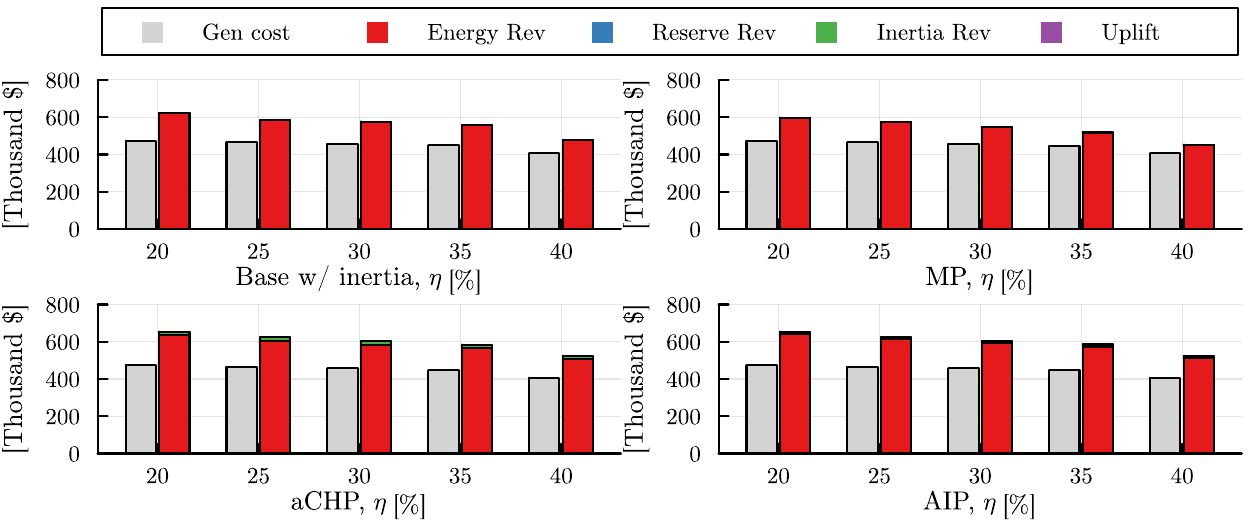}
        \label{fig:CBA_SG1}}
     \hfill
     \vspace{-3mm}
     \subfigure[Cost and revenue of SG 7 (mid-merit unit)]{
        \includegraphics[width=1\linewidth]{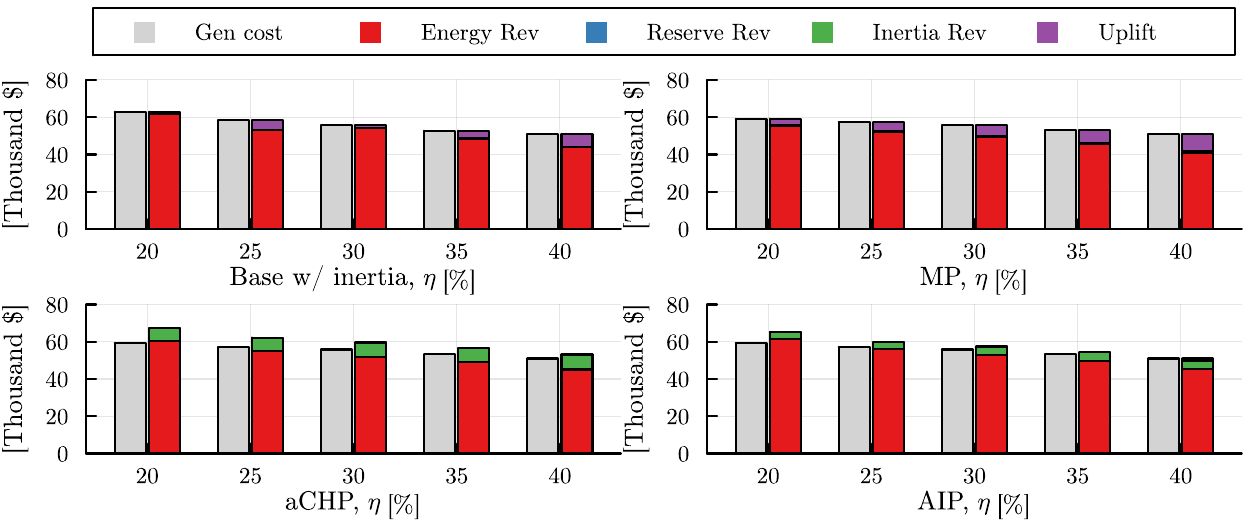}
        \label{fig:CBA_SG7}}
     \hfill
     \vspace{-3mm}
     \subfigure[Cost and revenue of SG 20 (peaking unit)]{
        \includegraphics[width=1\linewidth]{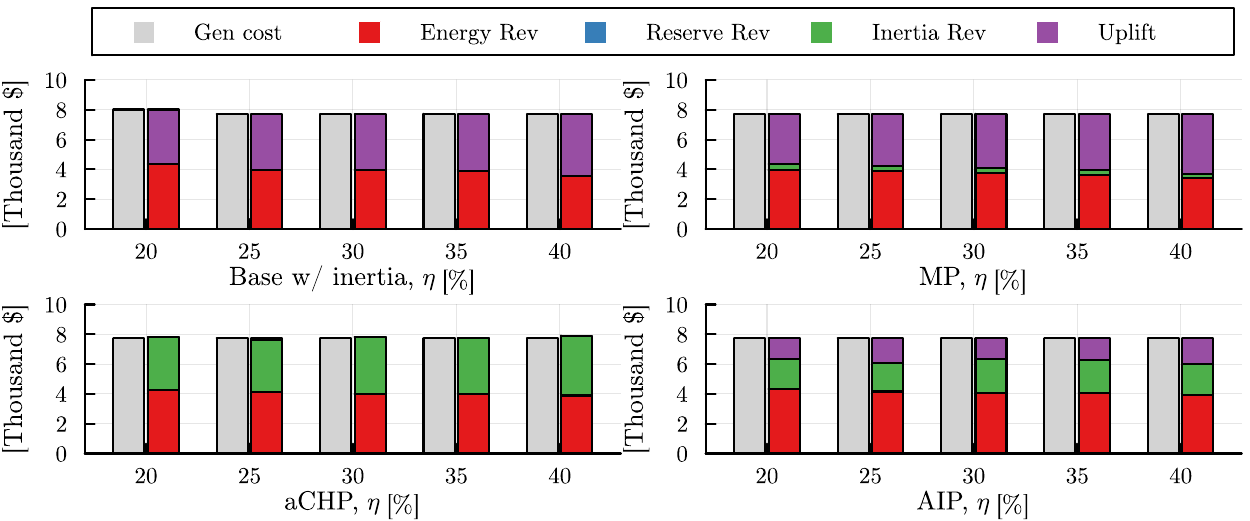}
        \label{fig:CBA_SG20}}
    \vspace{-3mm}
    \caption{\small Illustrating cost and revenue analysis of SG 1, SG 7, SG 20, and total renewable energy resources for each method.}
    \label{fig:CBA_SG}\vspace{-7mm}
\end{figure}


\subsubsection{Market Settlement Result}
Figs.~\ref{fig:CBA_SG1}–\ref{fig:CBA_SG20} present the cost and revenue structures for a base-load generator (SG~1), a mid-merit generator (SG~7), and a peaking generator (SG~20) under each pricing method. 
The results highlight that the economic impacts of different remuneration mechanisms are highly dependent on generator type.

\textbf{Base-load units} --  
SG~1 remains profitable across all methods, as its low marginal cost ensures continuous dispatch.
Uplift payments are unnecessary, since lower energy prices and reduced output occasionally limit total revenue in all scenarios.
All pricing methods enable energy and reserve prices to better reflect underlying operational costs; however, the impact is marginal for base-load units that are almost always online.

\textbf{Mid-merit units} --  
As $\eta$ increases, mid-merit generators become less profitable across all methods.
Because multiple SGs are committed for inertia provision, SG~7 often operates at its minimum generation limit while cheaper units set the market price. 
In MP, the average energy price (\$16.38/MWh) is below SG~7’s marginal cost at minimum output (\$17.58/MWh), resulting in revenue shortfalls. 
AIP mitigates this issue through the inclusion of average start-up and non-load costs in energy prices, though some uplift remains necessary.  
aCHP provides the strongest market signal among all methods for mid-merit units, as higher inertia prices compensate for reduced energy revenues, ensuring cost recovery without uplift.

\textbf{Peaking units} --  
The peaking generator is most affected by differences in pricing design. 
In Base w/ inertia, SG~20 is designated as a RMR unit and receives full compensation through uplift payments. 
In MP, the unit often operates only to provide inertia and cannot set prices due to minimum generation constraints, leading to severe revenue inadequacy (its marginal cost at minimum output, \$26.04/MWh, far exceeds the average market price of \$16.38/MWh).  
aCHP addresses this issue by introducing explicit inertia compensation, which provides an additional revenue stream that aligns payments with the value of inertia provision.  
AIP, while improving energy and reserve price adequacy, offers weaker inertia compensation than aCHP and still requires uplift in some cases.  

Overall, these results suggest that aCHP delivers the most economically consistent market signals across all generator types by jointly reflecting operational, commitment, and inertia-related costs. 
AIP provides moderate improvement relative to MP, particularly for mid-merit and peaking units, by incorporating fixed costs into price formation. 
In contrast, MP systematically undercompensates high-cost or infrequently dispatched units, highlighting the need for explicit inertia remuneration in future market designs.

\begin{figure}[t]%
    \centering
     \subfigure[$\eta = 15\%$]{%
        \includegraphics[width=0.475\linewidth]{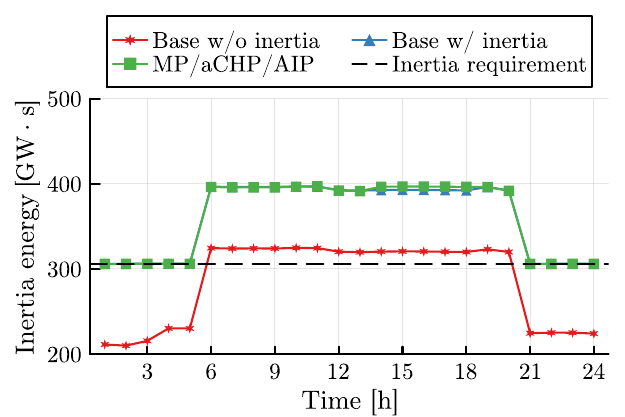}
        \label{fig:KPGSys_inertia_1}}
     \subfigure[$\eta = 20\%$]{
        \includegraphics[width=0.475\linewidth]{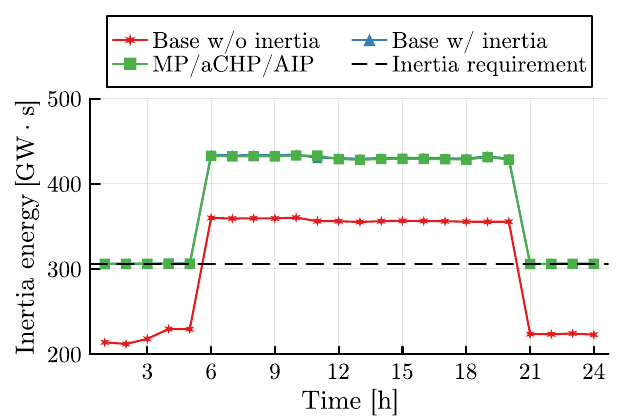}
        \label{fig:KPGSys_inertia_6}}
    \vspace{-3mm}\caption{\small \texttt{KPG193 Case} -- System inertia levels provided by all generation resources (SGs and RES) under different scenarios. The dashed line denotes the inertia requirement.
}\label{fig:KPGinertia_provision}
    \vspace{-7mm}
\end{figure}

\begin{figure}[t]%
    \centering
     \subfigure[At 01:00, $\eta = 15\%$]{%
        \includegraphics[width=0.48\linewidth]{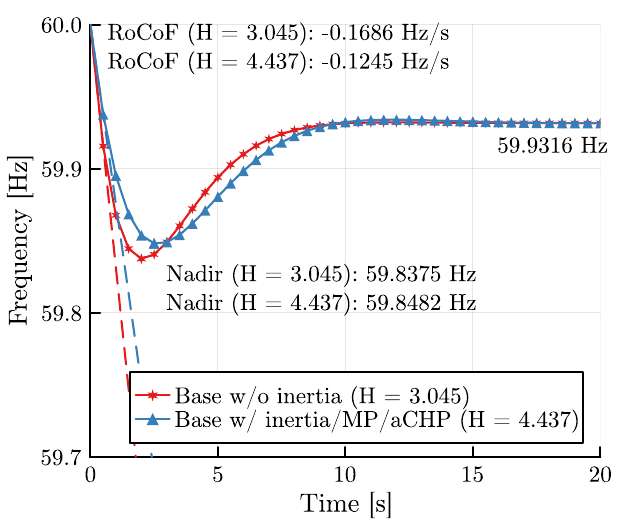}
        \label{fig:KPGSFR_H_15_min}}\hfill
     \subfigure[At 01:00, $\eta = 20\%$]{
        \includegraphics[width=0.48\linewidth]{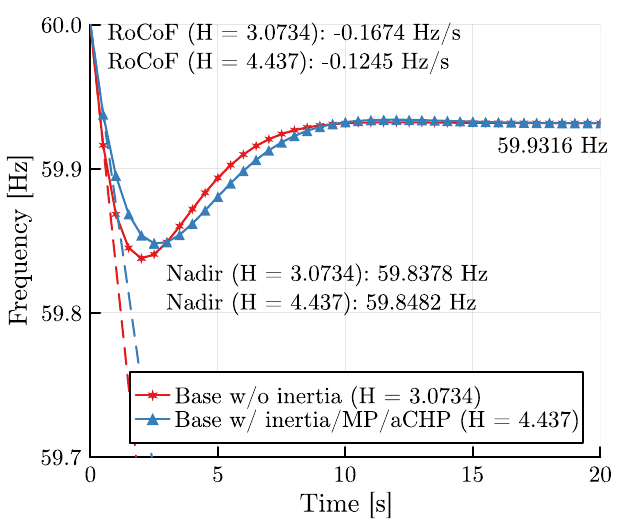}
        \label{fig:KPGSFR_H_20_min}}\vfill
    \vspace{-3mm}
     \subfigure[At 06:00, $\eta = 15\%$]{
        \includegraphics[width=0.48\linewidth]{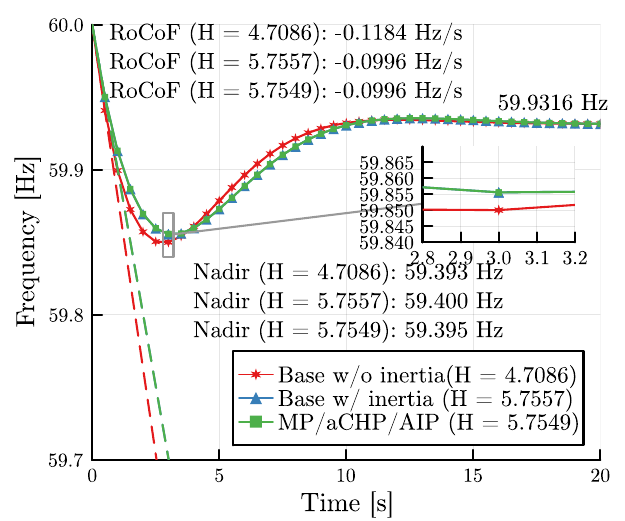}
        \label{fig:KPGSFR_H_15_max}}\hfill
     \subfigure[At 06:00, $\eta = 20\%$]{
        \includegraphics[width=0.48\linewidth]{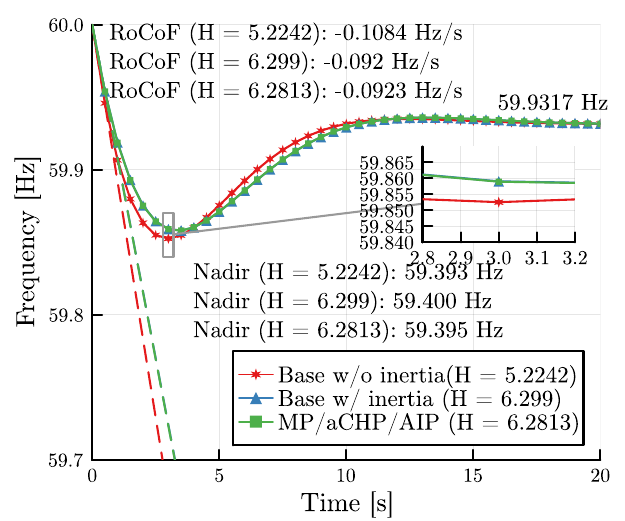}
        \label{fig:KPGSFR_H_20_max}}\vfill
    \vspace{-3mm}
    \caption{\small \texttt{KPG193 Case} --System frequency responses following a generation outage at different times (01:00 and 06:00).}
    \label{fig:KPGSFR}
    \vspace{-6mm}
\end{figure}

\begin{figure}[t]%
    \centering
     \subfigure[Base w/o inertia, $\eta = 15\%$]{%
        \includegraphics[width=0.48\linewidth]{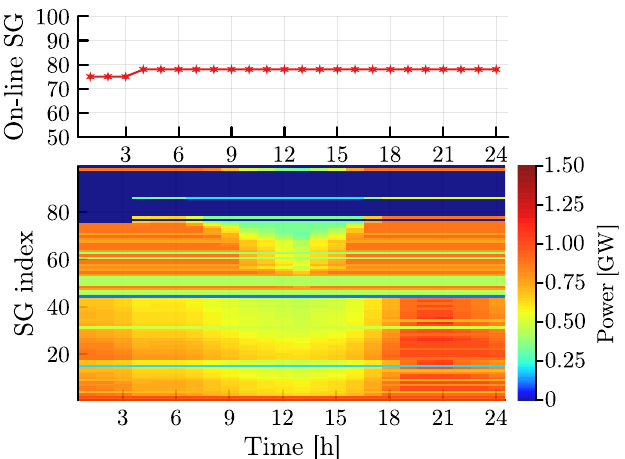}
        \label{fig:KPGgeneration_base1}}\hfill
     \subfigure[Base w/o inertia, $\eta = 20\%$]{
        \includegraphics[width=0.48\linewidth]{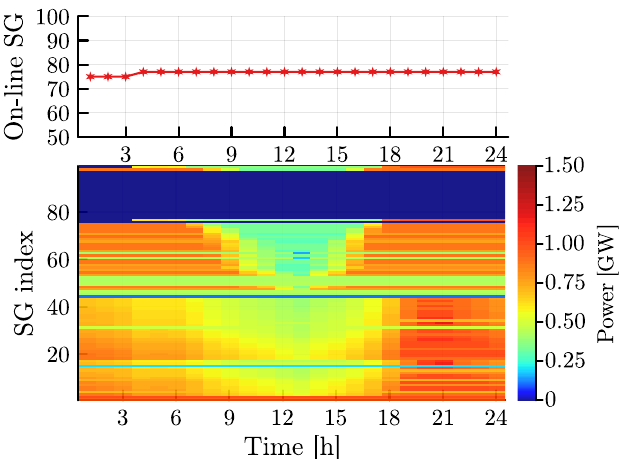}
        \label{fig:KPGgeneration_base6}}\vfill
        \vspace{-3mm}
     \subfigure[Base w/ inertia, $\eta = 15\%$]{
        \includegraphics[width=0.48\linewidth]{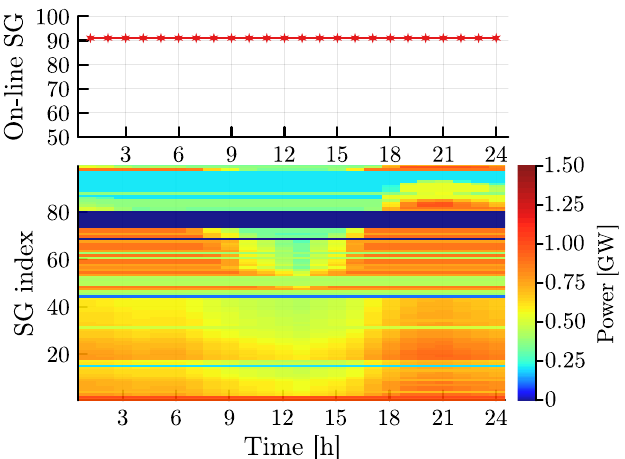}
        \label{fig:KPGgeneration_uplift1}}\hfill
     \subfigure[Base w/ inertia, $\eta = 20\%$]{
        \includegraphics[width=0.48\linewidth]{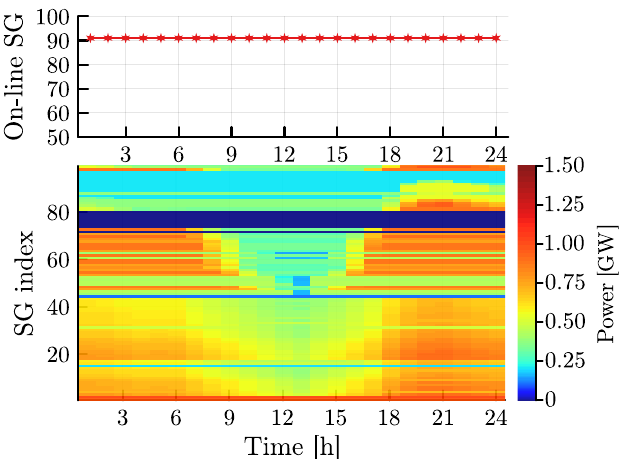}
        \label{fig:KPGgeneration_uplift6}}\vfill
        \vspace{-3mm}
     \subfigure[MP/aCHP/AIP, $\eta = 15\%$]{
        \includegraphics[width=0.48\linewidth]{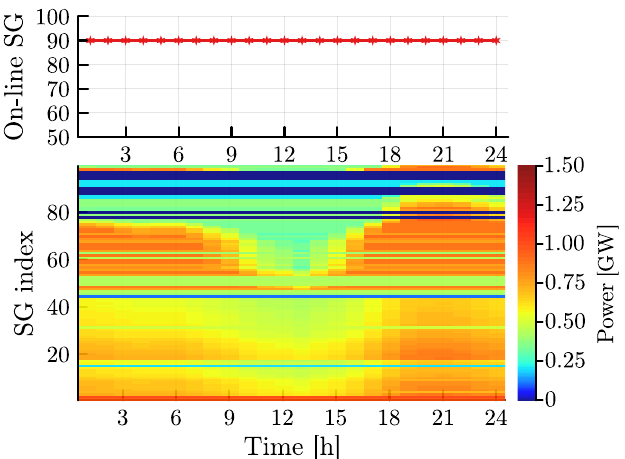}
        \label{fig:KPGgeneration_pricing1}}\hfill
     \subfigure[MP/aCHP/AIP, $\eta = 20\%$]{
        \includegraphics[width=0.48\linewidth]{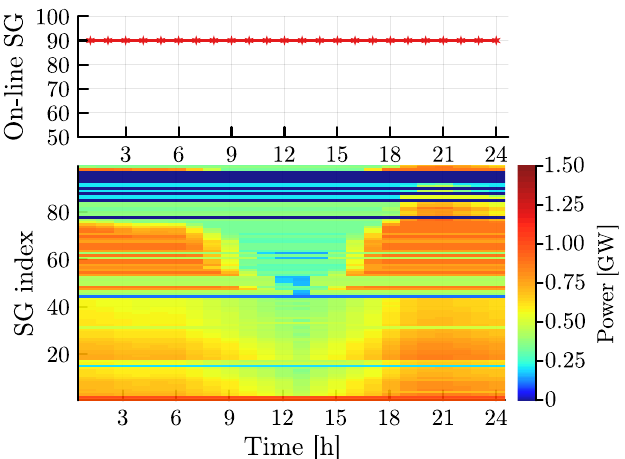}
        \label{fig:KPGgeneration_pricing6}}\vfill
    \vspace{-3mm}
    \caption{\small \texttt{KPG193 Case} -- Number of online SGs and their generation outputs under different scenarios and renewable penetration levels.}
    \label{fig:KPGgeneration}
    \vspace{-5mm}   
\end{figure}

\begin{figure}[t]
    \centering
     \subfigure[Energy price in MP, aCHP, and AIP]{
        \includegraphics[width=1.0\linewidth]{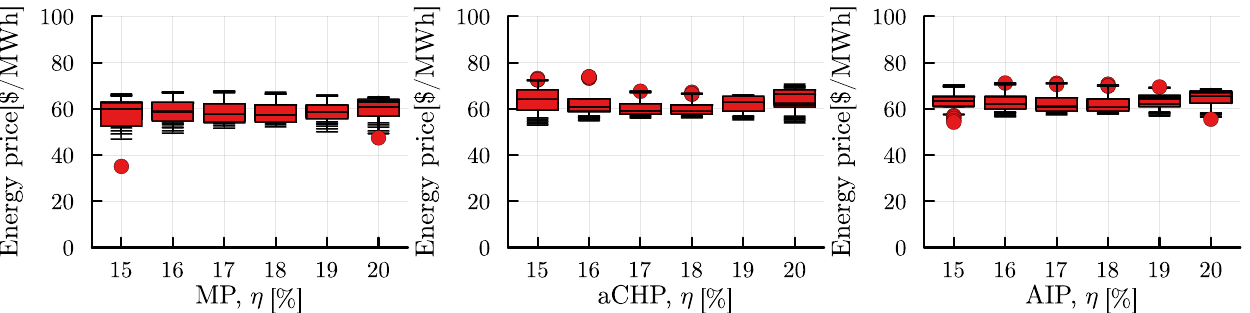}
        \label{fig:KPGEnergy_Price}}\vfill
        \vspace{-3mm}
     \subfigure[Reserve price in MP, aCHP, and AIP]{
        \includegraphics[width=1.0\linewidth]{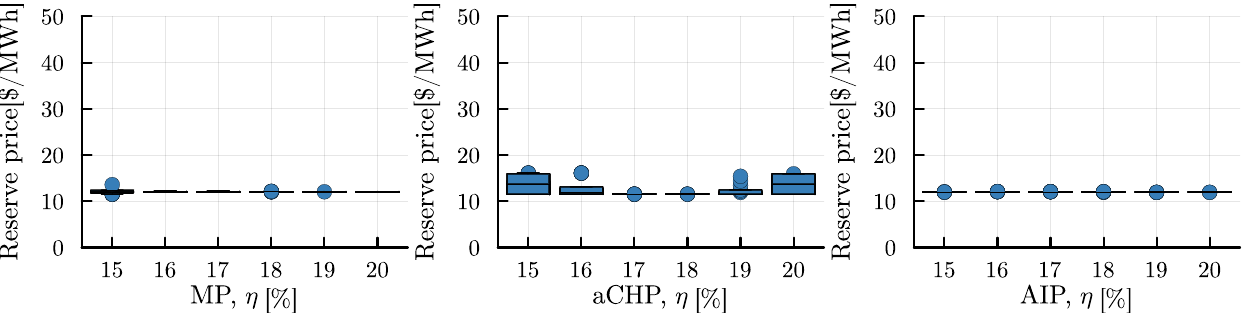}
        \label{fig:KPGReserve_Price}}\vfill 
        \vspace{-3mm}
     \subfigure[Inertia prices in MP, aCHP, and AIP]{
        \includegraphics[width=1.0\linewidth]{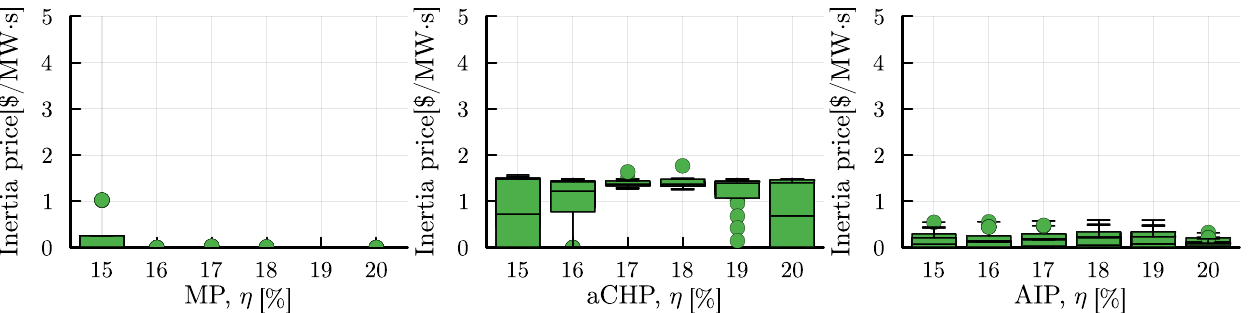}
        \label{fig:KPGInertia_Price}}\vfill     
    \vspace{-3mm}   
    \caption{\small \texttt{KPG193 Case} -- Box plots of 24-hour energy, reserve, and inertia prices.
    Each box represents the 25th–75th percentile range with the median line inside.}
    \label{fig:KPGprices}
    \vspace{-7mm}   
\end{figure}

\subsection{Extension to a Realistic Korean Test System}

To verify scalability and policy relevance, the proposed framework is applied to the realistic Korean power grid test system (KPG 193) with 122 thermal units (103,980 MW), 17 hydro units (7,200 MW), 193 PV units (23,743 MW), 94 WTs (1,626 MW) and 70 ESSs (1,400 MW) \cite{song2024kpg}.  
While the IEEE 118-bus case provided analytical validation under controlled conditions, the KPG 193 case demonstrates real-world applicability in a large, isolated system characterized by concentrated urban demand, rapid renewable growth, and stringent inertia requirements.  
This case highlights that the proposed inertia-aware unit-commitment (UC) and remuneration mechanisms remain computationally tractable and economically consistent at national-scale operation, while revealing several unique implications for Korea’s ongoing market-design discussions.

\subsubsection{Key similarities to the IEEE 118-bus case}  
\begin{figure}[t]
    \centering
     \subfigure[Cost and revenue of SG 1 (base-load unit)]{%
        \includegraphics[width=1\linewidth]{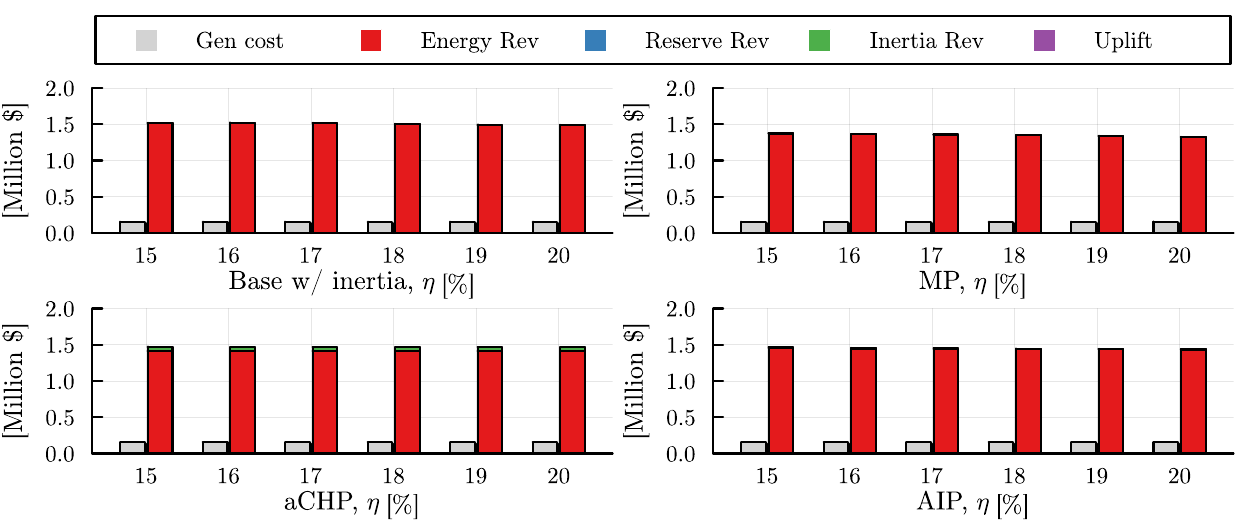}
        \label{fig:KPGCBA_SG1}}
     \hfill
     \vspace{-3mm}
     \subfigure[Cost and revenue of SG 30 (mid-merit unit)]{
        \includegraphics[width=1\linewidth]{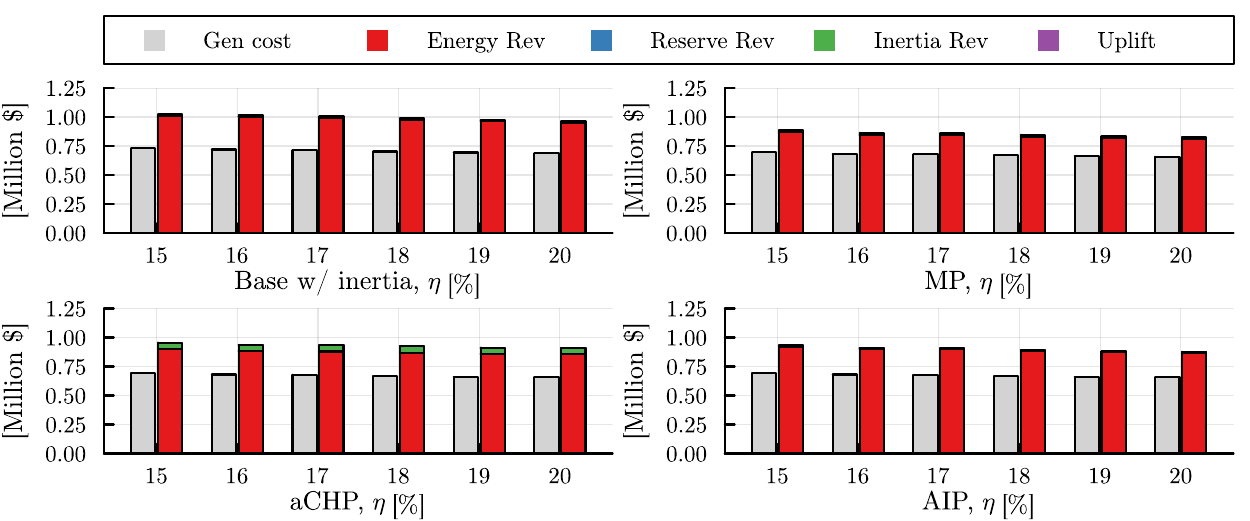}
        \label{fig:KPGCBA_SG30}}
    \hfill
    \vspace{-3mm}
    \subfigure[Cost and revenue of SG 90 (peaking unit)]{
        \includegraphics[width=1\linewidth]{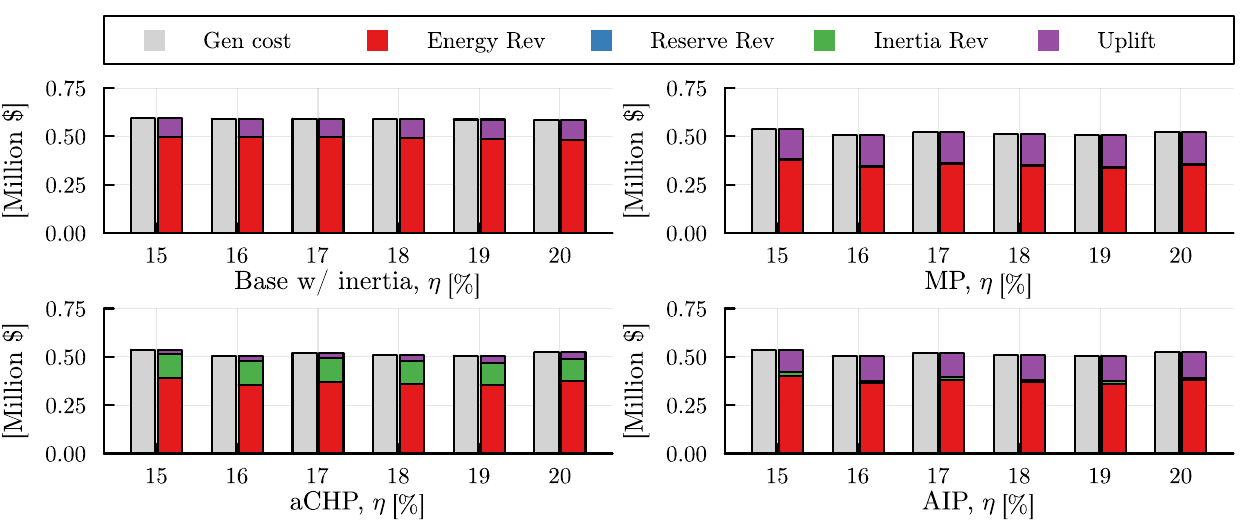}
        \label{fig:KPGCBA_SG90}}
    \vspace{-3mm}
    \caption{\small \texttt{KPG193 Case} -- Illustrating cost and revenue analysis of SG 1, SG 30, SG 90, and total renewable energy resources.}
    \label{fig:KPGCBA_SG}\vspace{-5mm}
\end{figure}

Overall operational patterns in Figs.~\ref{fig:KPGinertia_provision}--\ref{fig:KPGCBA_SG} are consistent: inertia-aware approaches (MP, aCHP, AIP) maintain frequency stability and reduce reliance on RMR contracts.  
The aCHP method again yields the highest inertia prices and lowest uplift payments, confirming its ability to ensure revenue adequacy without distorting market efficiency.  
AIP provides partial improvements by embedding average fixed costs, whereas MP continues to undervalue inertia and requires compensation through uplift.  

\subsubsection{Unique Characteristics of the Korean System}  
KPG 193 represents an isolated grid with no cross-border interconnections, which heightens the operational importance of maintaining adequate internal inertia because the system cannot rely on neighboring grids.
Many coal and LNG units in Korea operate near their minimum output levels, as shown in Fig.\ref{fig:KPGgeneration}, and therefore maintaining sufficient synchronous inertia further constrains overall system flexibility.
Inertia-aware operations commit roughly the same number of generators as the Base w/ inertia (RMR) case but replace high-cost RMR units with more economical mid-merit generators.  
This re-dispatch lowers total system costs while sustaining the required 306~GW·s inertia.  
Frequency analyses in Figs.~\ref{fig:KPGSFR_H_15_min}–\ref{fig:KPGSFR_H_20_max} indicate an approximately 35\% improvement in RoCoF during low-inertia hours.

\subsubsection{Market Outcomes and Policy Implications}  
In the KPG 193 system, inertia prices and their variance in Fig.~\ref{fig:KPGInertia_Price} remain higher than in the IEEE 118-bus case because of the high start-up costs of peaking units.
The aCHP method continues to produce the most transparent inertia price signal among the examined schemes, ensuring adequate compensation for peaking and mid-merit generators that supply inertia.
AIP moderates price volatility and provides more stable revenues for baseload and mid-merit units, but tends to under-compensate peaking units that contribute inertia under high renewable penetration.  
From a policy perspective, these results suggest that:
\begin{itemize}
    \item A dedicated inertia-pricing mechanism such as aCHP would enhance reliability in Korea’s future low-inertia grid by reducing dependence on RMR units.  
    \item Incorporating AIP can serve as a transitional measure to improve cost recovery and price stability before adopting convex-hull implementation.  
    \item Explicit inertia remuneration aligns economic incentives across various generators, particularly benefiting peaking units that contribute to grid operation but face declining operating hours under high-RES penetration.
\end{itemize}

Overall, the Korean case confirms that the proposed framework not only scales effectively but also provides actionable insights for designing a transparent and cost-reflective inertia market suited to Korea’s evolving power system.

\section{Conclusion}
This paper presents an inertia-aware CC-UC model incorporating time-coupling constraints for SGs and examines four remuneration methods—uplift, MP, aCHP, and AIP—applied to inertia-aware operation.
Simulation results show that the proposed model enhances frequency stability following generator contingencies.  
Among the remuneration methods, aCHP provided the most efficient market signal, ensuring revenue adequacy through inertia pricing while minimizing uplift payments.  
AIP offers moderate improvement by internalizing fixed costs, whereas MP required the highest uplift due to the absence of inertia valuation in its price formation.
The extended case study confirms the scalability and policy applicability of the proposed framework.  
In particular, the results suggest that adopting aCHP could reduce reliance on out-of-market RMR contracts and enhance investment incentives for peaking and mid-merit units.  
These findings provide practical guidance for designing transparent, cost-reflective inertia markets that support reliable operation in low-inertia systems.

\bibliographystyle{IEEEtran}
\bibliography{reference}

@misc{RES_capacity,
  title={{Renewables capacity statistics 2024}},
  author={{International Renewable Energy Agency (IRENA)}},
  date={2025}
}

@article{zhong2010synchronverters,
  title={Synchronverters: Inverters that mimic synchronous generators},
  author={Zhong, Qing-Chang and others},
  journal={IEEE transactions on industrial electronics},
  year={2010}
}

@article{8332112,
  author={IEEE},
  journal={IEEE Std 1547-2018}, 
  title={{IEEE} Standard for Interconnection and Interoperability of Distributed Energy Resources with Associated Electric Power Systems Interfaces}, 
  year={2018}
}

@techreport{CAISO3100,
  author      = {CAISO},
  title       = {{Establishing System Operating Limits for the Operations Horizon}},
  number      = {Operating Procedure No. 3100},
  year        = {2024},
}

@techreport{denholm2020inertia,
  author      = {Paul Denholm and others},
  title       = {Inertia and the Power Grid: A Guide Without the Spin},
  number      = {NREL/TP-6A20-73856},
  year        = {2020},
  month       = {May}
}

@techreport{ERCOT2021,
  author      = {{Electric Reliability Council of Texas (ERCOT)}},
  title       = {Inertia: Basic Concepts and Impacts on the ERCOT Grid},
  year        = {2018},
  month       = {April}
}

@techreport{ERCOT2024,
  author      = {{ERCOT}},
  title       = {One Pager: Reliability Must-Run (RMR) and Must-Run Alternative (MRA)},
  year        = {2024},
  month       = {March}
}

@techreport{CAISO2012BCR,
  author      = {{California Independent System Operator (CAISO)}},
  title       = {Draft Final Proposal: Bid Cost Recovery Mitigation Measures},
  year        = {2012},
  month       = {April}
}

@techreport{NGESO2023Stability,
  author      = {{National Grid Electricity System Operator (NG ESO)}},
  title       = {Stability Market Design: Final Outcomes},
  institution = {National Grid ESO},
  number      = {NIA2\_NGESO005},
  year        = {2021},
  month       = {December}
}

@techreport{neso2021tpr,
  title        = {Technical Performance Requirements Version 2},
  author       = {{National Energy System Operator (NESO)}},
  year         = {2021},
  month        = dec
}

@techreport{Eirgrid,
  author      = {{Single Electricity Market Committee}},
  title       = {Contractual Arrangements for Low Carbon Inertia Services (LCIS) – Decision Paper},
  institution = {Single Electricity Market Committee (SEMC)},
  number      = {SEM-23-064},
  year        = {2023},
  month       = {Sep.}
}

@techreport{eirgrid2022lcis,
  title        = {Consultation on Low Carbon Inertia Service (LCIS) Competitive Procurement},
  author       = {{EirGrid and SONI}},
  institution  = {EirGrid Group},
  year         = {2022},
  month        = jun
}

@techreport{WEMRules2025,
  author      = {{Government of Western Australia}},
  title       = {Wholesale Electricity Market Rules},
  year        = {2025},
}

@techreport{AEMO2024FCESS,
  author      = {{Australian Energy Market Operator}},
  title       = {WEM Procedure: Frequency Co-Optimised Essential System Services Accreditation},
  year        = {2024},
}

@article{doherty2005frequency,
  title={Frequency control in competitive electricity market dispatch},
  author={Doherty, Ronan and others},
  journal={IEEE Transactions on Power Systems},
  year={2005},
  publisher={IEEE}
}

@article{hu2023inertia,
  title={Inertia Market: Mechanism Design and its Impact on Generation Mix},
  author={Hu, Jingwei and others},
  journal={J. Mod. Power Syst. Clean Energy},
  year={2023},
  publisher={SGEPRI}
}

@article{paturet2020stochastic,
  title={Stochastic unit commitment in low-inertia grids},
  author={Paturet, Matthieu and others},
  journal={IEEE Transactions on Power Systems},
  year={2020},
  publisher={IEEE}
}

@article{badesa2019simultaneous,
  title={Simultaneous scheduling of multiple frequency services in stochastic unit commitment},
  author={Badesa, Luis and others},
  journal={IEEE Transactions on Power Systems},
  volume={34},
  number={5},
  pages={3858--3868},
  year={2019},
  publisher={IEEE}
}

@article{liang2022inertia,
  title={Inertia pricing in stochastic electricity markets},
  author={Liang, Zhirui and others},
  journal={IEEE Transactions on Power Systems},
  volume={38},
  number={3},
  pages={2071--2084},
  year={2022},
  publisher={IEEE}
}

@article{paturet2020economic,
  title={Economic valuation and pricing of inertia in inverter-dominated power systems},
  author={Paturet, Matthieu and others},
  journal={arXiv preprint arXiv:2005.11029},
  year={2020}
}

@article{qiu2024market,
  title={Market design for ancillary service provisions of inertia and frequency response via virtual power plants: A non-convex bi-level optimisation approach},
  author={Qiu, Dawei and others},
  journal={Applied Energy},
  volume={361},
  pages={122929},
  year={2024},
  publisher={Elsevier}
}

@article{lu2024convex,
  title={Convex-hull Pricing of Ancillary Services for Power System Frequency Regulation with Renewables and Carbon-Capture-Utilization-and-Storage Systems},
  author={Lu, Zelong and others},
  journal={IEEE Transactions on Power Systems},
  year={2024},
  publisher={IEEE}
}

@misc{o2019essays,
  title={Essays on average incremental cost pricing for independent system operators},
  author={O’Neill, Richard and others},
  year={2019},
  publisher={Working paper, Washington, DC}
}

@article{hua2016convex,
  title={A convex primal formulation for convex hull pricing},
  author={Hua, Bowen and Baldick, Ross},
  journal={IEEE Transactions on Power Systems},
  year={2016},
  publisher={IEEE}
}

@article{wang2016commitment,
  title={Commitment cost allocation of fast-start units for approximate extended locational marginal prices},
  author={Wang, Congcong and others},
  journal={IEEE Transactions on Power Systems},
  volume={31},
  number={6},
  pages={4176--4184},
  year={2016},
  publisher={IEEE}
}

@article{dvorkin2019chance,
  title={A chance-constrained stochastic electricity market},
  author={Dvorkin, Yury},
  journal={IEEE Transactions on Power Systems},
  volume={35},
  number={4},
  pages={2993--3003},
  year={2019},
  publisher={IEEE}
}

@article{kargarian2015chance,
  title={Chance-constrained system of systems based operation of power systems},
  author={Kargarian, Amin and others},
  journal={IEEE Transactions on Power systems},
  volume={31},
  number={5},
  pages={3404--3413},
  year={2015},
  publisher={IEEE}
}

@article{dreidy2017inertia,
  title={Inertia response and frequency control techniques for renewable energy sources: A review},
  author={Dreidy, Mohammad and others},
  journal={Renewable and sustainable energy reviews},
  volume={69},
  pages={144--155},
  year={2017},
  publisher={Elsevier}
}

@ARTICLE{7904729,
  author={Peña, Ivonne and others},
  journal={IEEE Transactions on Power Systems}, 
  title={An Extended IEEE 118-Bus Test System With High Renewable Penetration}, 
  year={2018},
}

@article{Carrion,
  author={Carrion, M. and Arroyo, J.M.},
  journal={IEEE Transactions on Power Systems}, 
  title={A computationally efficient mixed-integer linear formulation for the thermal unit commitment problem}, 
  year={2006},
  volume={21},
  number={3},
  pages={1371-1378},
  doi={10.1109/TPWRS.2006.876672}}

@misc{Caiso,
    title={{CAISO} generation mix in 2023},
    author={CAISO},
    year = {2024}, 
}

@article{villena2024assessment,
  title={Assessment of the synthetic inertial response of an actual solar PV power plant},
  author={Villena-Ruiz, Raquel and others},
  journal={International Journal of Electrical Power \& Energy Systems},
  volume={157},
  pages={109875},
  year={2024},
  publisher={Elsevier}
}

@article{anderson1990low,
  title={A low-order system frequency response model},
  author={Anderson, Philip M and Mirheydar, Mahmood},
  journal={IEEE Transactions on power systems},
  year={1990},
  publisher={IEEE}
}

@article{fernandez2019power,
  title={Power systems with high renewable energy sources: A review of inertia and frequency control strategies over time},
  author={Fern{\'a}ndez-Guillam{\'o}n, Ana and others},
  journal={Renewable and Sustainable Energy Reviews},
  year={2019},
  publisher={Elsevier}
}

@article{song2024kpg,
  title={KPG 193: A Synthetic Korean Power Grid Test System for Decarbonization Studies},
  author={Song, Geonho and Kim, Jip},
  journal={arXiv:2411.14756},
  year={2024}
}
\end{document}